\documentclass{jfm}

\usepackage[utf8]{inputenc}
\usepackage[T1]{fontenc}

\usepackage{graphicx}
\usepackage{natbib}
\usepackage{amsmath}
\usepackage{amssymb}
\usepackage{hyperref}
\usepackage{nicefrac}
\usepackage{xcolor}
\usepackage[english=british]{csquotes}
\usepackage{nicefrac}

\DeclareMathAlphabet{\mathbbold}{U}{bbold}{m}{n}
\renewcommand{\vec}[1]{\boldsymbol{#1}}
\newcommand{\mat}[1]{\boldsymbol{\mathrm{#1}}}
\newcommand{\zero}{\mathbbold{0}}
\newcommand{\one}{\mathbbold{1}}
\newcommand{\mX}{\mat{X}}
\newcommand{\mS}{\mat{S}}
\newcommand{\mT}{\mat{T}}
\newcommand{\mC}{\mat{C}}
\newcommand{\mL}{\mat{\Lambda}}
\newcommand{\mLabs}{\mL_+}
\newcommand{\mLs}{\mL_s}
\newcommand{\mP}{\mat{\Phi}}
\newcommand{\mXi}{\mat{\Xi}}

\renewcommand{\Pr}{{\textit{Pr}}}
\newcommand{\Ra}{{\textit{Ra}}}
\newcommand{\Nu}{{\textit{Nu}}}
\newcommand{\Nuc}{{\textit{Nu}}^c}

\newcommand{\R}{\ensuremath{\mathbb{R}}}

\DeclareMathOperator{\diag}{diag}

\DeclareMathOperator{\sgn}{sgn}

\title[Describing the Heat Transport of Turbulent RB Convection by POD methods]{Describing the Heat Transport of Turbulent Rayleigh--Bénard Convection by POD methods}

\author[J.~L{\"u}lff]{Johannes L{\"u}lff$^1$\thanks{Email address for correspondence: johannes.luelff@uni-muenster.de}}

\affiliation{$^1$Institute for Theoretical Physics, University of M{\"u}nster, Wilhelm-Klemm-Str.~9, 48149 M{\"u}nster, Germany}

\begin{document}

% \tableofcontents
% \newpage

\maketitle

\begin{abstract}
	Rayleigh--Bénard convection, which is the buoyancy-induced motion of a fluid enclosed between two horizontal plates, is an idealised setup to study thermal convection.
	We analyse the modes that transport the most heat between the plates by computing the proper orthogonal decomposition (POD) of numerical data.
	Instead of the usual POD ansatz of finding modes that describe the energy best, we propose a method that is optimal in describing the heat transport.
	Thereby, we can determine the modes with the major influence on the heat transport and the coherent structures in the convection cell.
	We also show that in lower-dimensional projections of numerical convection data, the newly developed modes perform consistently better than the standard modes.
	We then use this method to analyse the main modes of three-dimensional convection in a cylindrical vessel as well as two-dimensional convection with varying Rayleigh number and varying aspect ratio.
\end{abstract}

\section{Introduction}
% \begin{itemize}
% 	\item RB convection: erratic, hard to handle analytically
% 	\item Nevertheless: coherent structures appear, e.g. plumes or large-scale circulation
% 	\item Our focus: main modes of convection and their heat transport
% 	\item Technique for obtaining coherent structures: proper orthogonal decomposition (POD)
% 	\item POD: method from data analysis, used to obtain a basis from a data ensemble of snapshots that gives the \enquote{best} representation of the data
% 	\item Application of POD to RBC:\citep{sirovich87qam, sirovich90pfa, lumley97pof, bailoncuba10jfm, bailoncuba11pof}
% 	\item \enquote{Best} means: modes capture the most $L^2$-norm of the data set, i.e. most \enquote{generalized} energy
% 	\item Our new ansatz proposed in this paper: Instead of the most generalized energy, we extend the method to give the most heat transport, measured in terms of the Nusselt number
% \end{itemize}

In Rayleigh--Bénard convection, the motion of a fluid enclosed between two horizontal plates is driven by injecting heat at the hot bottom and removing it at the cold top, thus inducing a heat transport between the plates.
This simple principle is at the core of many flows that occur in nature, with examples including the motion of the oceans and the atmosphere as well the tectonics inside the Earth and the convection patterns observed at the surface of the Sun.
For an overview of recent findings in Rayleigh--Bénard research, see the reviews by \citet{ahlers09rmp}, \citet{lohse10rfm} and \citet{chilla12epj}.

When the temperature difference (measured in terms of the Rayleigh number $\Ra$) is small, the system displays stable laminar convection.
It is long known that in this case the analytical solution of the underlying equations of motion \citep{oberbeck79apc,boussinesq03book} is possible, and patterns in the form of coherent convection rolls are obtained \citep{rayleigh16phm}.
On the other hand, convective flows that occur in nature generally have a high Rayleigh number and are very turbulent and erratic, which prevents a direct analytical solution.
Nevertheless, amidst the small-scale turbulence, coherent structures like a large-scale circulation emerge from the system.
We want to focus on these main modes of convection and understand how the irregular flow fields of turbulent convection are composed from these building blocks and how they contribute to the heat transport through the system.

To this end, the technique of \emph{proper orthogonal decomposition} (POD) is often used to obtain a hierarchy of orthogonal basis modes (the so-called POD modes) that build the turbulent fields in a top-down fashion, starting from the strongest large-scale structures and continuing with higher-order modes until the finest scales of turbulence filaments are reached.
This technique, borrowed from the field of data analysis, where it is known as the Karhunen--Loève transformation \citep{karhunen46asf,loeve45cra} or the principal component analysis \citep{pearson01pms}, is able to extract the main modes from a data set.
In a fluid dynamical context, a data set is a series of flow fields at different time instants, and the POD then gives the \enquote{best} description of the data set.
The method is therefore purely data-driven as it does not rely on any underlying equations of motion, and it may be applied to experimentally or numerically obtained data sets.
The concept of POD was mainly introduced to the field of fluid dynamics by \citet{sirovich87qam} and has since been used to describe coherent structures in e.g. channel flows \citep{sirovich87qam2}, flat plate boundary layers \citep{rempfer94jfm1} and Rayleigh--Bénard convection \citep{bailoncuba10jfm}.
Also, the dynamical behaviour of the POD modes has been analysed by means of Galerkin projections by \citet{rempfer94jfm2}, \citet{rowley04phd} and \citet{bailoncuba11pof}.
For an introduction into POD aimed at an application to fluid dynamics, we refer to \citet{holmes96book} and \citet{smith05ndy}.

The property of POD that it gives the \enquote{best} set of basis modes means that the POD modes capture a maximal amount of (generalised) energy from the data set.
But, although mathematically sound, the generalised energy is in this case not a physically meaningful quantity.
The goal of the paper therefore is to adapt the POD technique to the Rayleigh--Bénard system such that it instead gives the best description of the convective heat transport from hot bottom to cold top plate, measured in terms of the Nusselt number.
The understanding and scaling of the Nusselt number is one of the key questions in Rayleigh--Bénard research \citep{ahlers09rmp}, and its interplay with the structure of the large-scale circulation is at the core of many theories that try to predict the Nusselt number \emph{a priori} \citep{castaing89jfm,grossmann00jfm, grossmann01prl}.

The remainder of this paper is structured as follows:
In section~\ref{sec:theory} we give a brief introduction into the mathematics behind POD and then show how we adapt it to measure the convective heat transport.
Section~\ref{sec:results} reports on the results of applying the new method to data sets obtained from direct numerical simulation:
First we benchmark the new method by comparing it to the standard approach in section~\ref{sec:results_2d}. We then utilise the new technique to analyse the main structures and their heat transport of three-dimensional cylindrical convection (section~\ref{sec:results_3d}) and two-dimensional convection with varying Rayleigh number (section~\ref{sec:results_rascan}) and aspect ratio (section~\ref{sec:results_gammascan}).

\section{Theory}\label{sec:theory}
In this section we will start with recounting the ideas behind the POD technique and the method of snapshots, first introduced by \citep{sirovich87qam1}, in a matrix formulation (sections \ref{sec:variational_problem}--\ref{sec:mos}); the reader is also referred to \citet{rowley04phd} and \cite{smith05ndy} for a more in-depth introduction into the mathematics.
In section \ref{sec:new_approach} we then introduce our modified approach that is able the describe the convective heat transport.

\subsection{Variational Problem}\label{sec:variational_problem}
We start the derivation from a data ensemble, i.e. a set $\{\vec{q}(\vec{x})\}$ of snapshots or state vectors. % that may have been obtained by numerical or experimental means.
For two-dimensional Rayleigh--Bénard convection, a snapshot $\vec{q}(\vec{x})$ consists of the flow fields of temperature and velocity.
Due to incompressibility, the horizontal velocity is already determined by the vertical velocity $u_z$, and thus the state vector
\begin{equation}\label{eq:snapshot_continuous}
	\vec{q}(\vec{x}) = \left(\begin{matrix}T(\vec{x})\\u_z(\vec{x})\end{matrix}\right)\;,\;\vec{q}:\Omega\subset\R^2\rightarrow\R^2
\end{equation}
fully characterises the flow configuration.
Here, $\Omega\subset\R^2$ is the fluid domain with volume $V$.
We will consider the temperature field to be non-dimensionalised with respect to the temperature difference between the plates, and the velocity is made non-dimensional by describing it in terms of the velocity of diffusion.

The goal is to find a set of orthonormal basis modes $\{\vec{\phi}(\vec{x})\}$ (\enquote{POD modes}) that give the best description of the data set in the sense that among all possible sets of basis modes, the POD modes minimise the $L^2$-error between a lower-dimensional projection and the full data set.
In other words, the POD modes maximise the average $L^2$-norm of the projection onto the POD modes:
\begin{equation}\label{eq:variational_problem}
	\bigl\langle|(\vec{q},\vec{\phi})|^2\bigr\rangle\rightarrow\text{max.}\quad\text{under normalisation constraint}\quad(\vec{\phi},\vec{\phi})=1
\end{equation}
Here, $(\cdot,\cdot)$ is a general scalar product of two fields, e.g. $(\vec{f},\vec{g}) = \int_\Omega\!\mathrm{d}^2x\,\vec{f}(\vec{x})^\dagger\mS\vec{g}(\vec{x})$ with $\mS\in\R^{2\times 2}$ positive definite and Hermitian; the usual and most simple case $\mS=\one$ results in the $L^2$-norm $(\vec{q},\vec{q})=\int_\Omega\!\mathrm{d}^2x\,\bigl(T^2+u_z^2\bigr)$ of a snapshot, which we call its generalised energy.
The dagger $(\cdot)^\dagger$ indicates the conjugate transpose; although we will only consider real valued data sets, the method can also be applied to complex fields, and therefore we will keep the description as general as possible (this is also the reason why we e.g. demand $\mS$ to be Hermitian instead of symmetric).
The average $\langle\cdot\rangle$ is calculated over the ensemble $\{\vec{q}(\vec{x})\}$ and in our case is considered as a temporal average by assuming ergodicity.

According to variational calculus, the maximisation of the functional \eqref{eq:variational_problem} is equivalent to solving the eigenvalue equation
\begin{equation}\label{eq:evp}
	\int_\Omega\!\mathrm{d^2}x'\,\bigl\langle\vec{q}(\vec{x})\otimes\vec{q}(\vec{x}')\bigr\rangle\mS\vec{\phi}(\vec{x'}) = \lambda\vec{\phi}(\vec{x})
\end{equation}
where $\vec{q}(\vec{x})\otimes\vec{q}(\vec{x}') = \vec{q}(\vec{x})\vec{q}(\vec{x}')^\dagger$ is the outer product, and $\lambda$ stems from a Lagrange multiplier due to normalisation (also cf.~\citet{rowley04phd}).
Thus, the POD modes $\vec{\varphi}(\vec{x})$ are the eigenfunctions of the temporally averaged spatial covariance kernel (or matrix) $\bigl\langle\vec{q}(\vec{x})\otimes\vec{q}(\vec{x}')\bigr\rangle$, and the eigenvalues $\lambda$ give the average projection of the data set onto the corresponding eigenfunction, i.e. $\lambda=\bigl\langle|(\vec{q},\vec{\phi})|^2\bigr\rangle$, which is the generalised energy represented by mode $\vec{\phi}$.
The covariance kernel has Hermitian symmetry and the eigenvalues are real and positive, and the eigenfunctions are orthonormal with respect to the scalar product induced by $\mS$.

\subsection{Matrix Formulation for Discrete Data}
Usually the members of the data ensemble are not continuous functions as implied by \eqref{eq:snapshot_continuous}, but instead the data is obtained by experimental or numerical measurements, and each snapshot is represented by $N_x$ samples.
For the case of 2D convection data obtained from DNS, the temperature and vertical velocity fields are sampled at $n_x$ grid points, and a snapshot is represented by the vector
\begin{equation}\label{eq:snapshots}
	\vec{q}=(T_1,\dots,T_{n_x},u_{z,1},\dots,u_{z,n_x})^T\in\R^{N_x}
\end{equation}
with $N_x=2\,n_x$ spatial degrees of freedom.
The data ensemble consists of $N_t$ snapshots (counted by an upper index), and the whole ensemble can be represented as the data matrix
\begin{equation}
	\mat{X} = \left(\vec{q}^1\,\vec{q}^2\,\cdots\,\vec{q}^{N_t}\right)\in\R^{N_x\times N_t}.
\end{equation}
We will assume that $\mX$ has full rank; otherwise it contains redundant information (e.g., the same snapshot twice) that can easily be left out of the calculation.

The discrete version of the scalar product is induced by the positive definite Hermitian matrix $\mS\in\R^{N_x\times N_x}$, which in general is diagonal and accounts for the spatial weights of the grid points.
For the generalised energy as in section \ref{sec:variational_problem} it reads
\begin{equation}\label{eq:mS_E}
	\mS = \left(\begin{matrix}\mat{W} & \zero \\ \zero & \mat{W} \end{matrix} \right)
\end{equation}
with the diagonal matrix $\mat{W}\in\R^{n_x\times n_x}$ containing the spatial weights.
The temporal averaging is carried by the Hermitian matrix $\mT\in\R^{N_t\times N_t}$ which is also positive definite and usually contains the temporal weights on its diagonal; the most simple case would be $\mT=\frac{1}{N_t}\one$ for uncorrelated snapshots or equidistant sampling in time.

With these prerequisites, the spatial covariance matrix becomes $\mX\mT\mX^\dagger\in\R^{N_x\times N_x}$, and the discrete formulation of the eigenvalue problem \eqref{eq:evp} reads
\begin{equation}\label{eq:evp_discrete}
	\mX\mT\mX^\dagger\mS\mP=\mP\mL.
\end{equation}
Here, the matrix $\mP\in\R^{N_x\times N_x}$ contains the POD modes $\vec{\phi}^i$ as columns, and $\mL\in\R^{N_x\times N_x}$ is diagonal containing the corresponding eigenvalues $\lambda_i$.
As $\mX\mT\mX^\dagger\mS$ is self-adjoined with respect to the scalar product induced by $\mS$, its eigenvectors can be chosen to be orthonormal under the same scalar product, i.e.
\begin{align}\label{eq:orthonormal_pod}
	\mP^\dagger\mS\mP=\one.
\end{align}
Also, for positive definite $\mS$ and $\mT$, the eigenvalues are non-negative, which is in compliance with their interpretation as generalised energies or $L^2$-norms.

\subsection{Method of Snapshots}\label{sec:mos}
Although \eqref{eq:evp_discrete} (the \enquote{direct method}) uniquely defines the POD modes, the calculation becomes impossible for the application we are aiming at.
This is because the data set usually contains much more spatial degrees of freedom than temporal snapshots, i.e. $N_x\gg N_t$, and therefore the spatial covariance matrix $\mX\mT\mX^\dagger$ becomes vastly too large to handle.
For example, $N_x=\mathcal{O}(10^7)$ and $N_t=\mathcal{O}(10^3)$ for the cylindrical data used later on, and the spatial covariance matrix would consume roughly $6\times10^6$ gigabyte of data.

However, for $N_x>N_t$ the rank of the spatial covariance matrix is $N_t$, and thus, $N_t$ non-zero eigenvalues and therefore only $N_t$ physically meaningful POD modes that actually contain energy can be obtained.
Thus, the POD modes span the same $N_t$-dimensional subspace of $\R^{N_x}$ as the snapshots, and accordingly, the POD modes can be constructed as linear combinations of the snapshots (remember that we assumed the snapshots to be linearly independent):
\begin{equation}\label{eq:modes_mos}
	\mP=\mX\mT\mC\mL^{-\nicefrac12}
\end{equation}
with $\mC\in\R^{N_t\times N_t}$ determining the linear combinations and $\mT$ and $\mL$ for normalisation reasons (choosing this special linear combination becomes reasonable shortly).
Thus, the calculation of the POD modes boils down to obtaining the matrix $\mC$, and inserting the linear combination \eqref{eq:modes_mos} into the eigenvalue problem \eqref{eq:evp_discrete} of the direct method readily yields
\begin{equation}\label{eq:evp_mos}
	\mX^\dagger\mS\mX\mT\mC=\mC\mL.
\end{equation}
The matrix $\mC$ contains the eigenvectors of $\mX^\dagger\mS\mX\mT$ as columns, where $\mX^\dagger\mS\mX\in\R^{N_t\times N_t}$ is the spatially averaged temporal covariance matrix.
The eigenvalues on the diagonal of $\mL$ are the same as obtained from the direct method \eqref{eq:evp_discrete}, because the matrices $\mX^\dagger\mS\mX\mT$ and $\mX\mT\mX^\dagger\mS$ share the same non-zero eigenvalues.
Also, $\mX^\dagger\mS\mX\mT$ is self-adjoint with respect to $\mT$, and thus, 
\begin{align}\label{eq:orthonorm_C}
	\mC^\dagger\mT\mC = \one,
\end{align}
cf. \eqref{eq:orthonormal_pod}.
With this, the orthonormality of the POD modes \eqref{eq:modes_mos} follows:
\begin{subequations}
\begin{align}
	\mP^\dagger\mS\mP &= \mL^{-\nicefrac12}\mC^\dagger\mT^\dagger\underbrace{\mX^\dagger\mS\mX\mT\mC}_{=\mC\mL}\mL^{-\nicefrac12}\\
		&= \mL^{-\nicefrac12}\underbrace{\mC^\dagger\mT^\dagger\mC}_{=\one}\mL\mL^{-\nicefrac12} = \one
\end{align}
\end{subequations}

Equations \eqref{eq:modes_mos} and \eqref{eq:evp_mos} together constitute the so-called method of snapshots, which yields the same physically meaningful POD modes spanning the data ensemble as the direct method.
Since it relies on the eigenvalue problem of the $N_t\times N_t$ temporal instead of the $N_x\times N_x$ spatial covariance matrix, this way of calculating the POD modes is preferable whenever there are more grid points than snapshots, as will virtually always be the case in the application we are aiming at.
Therefore, we only consider the method of snapshots from now on.

\subsection{Transformation into POD Basis and Lower-Dimensional Projection}\label{sec:transformation}
The POD modes as in \eqref{eq:modes_mos} form an orthonormal basis of the linear hull of the data set.
This basis may be used for a transformation of the data set into the POD basis or for a lower-dimensional projection of the data.

To transform the data into the POD basis, the amplitudes $\mXi\in\R^{N_t\times N_t}$ of the POD modes have to be found such that
\begin{align}\label{eq:transformation}
	\mX = \mP\mXi.
\end{align}
The $i$-th row of $\mXi$ gives the time series of mode $\vec{\phi}^i$, and the $j$-th column is the partition of snapshot $\vec{q}^j$.
In principle, the amplitudes may be found by projecting the data set onto the POD basis; instead of this direct calculation involving the big $N_x\times N_t$ matrices $\mX$ and $\mP$, this information is more easily obtained from the eigenvalue problem of the temporal covariance matrix:
\begin{align}\label{eq:prefactors}
	\mXi = \mP^\dagger\mS\mX=\mL^{\nicefrac12}\mC^\dagger
\end{align}
Here, the relations \eqref{eq:modes_mos}, \eqref{eq:evp_mos} and \eqref{eq:orthonorm_C} have been used.

A lower-dimensional projection of the data set is obtained by considering only a number $N_p<N_t$ of POD modes and corresponding amplitudes in \eqref{eq:transformation}.
Since the ordering of the modes in $\mP$ is arbitrary, we will assume a projection onto the first $N_p$ modes, which will usually be the most energetic ones corresponding to the biggest eigenvalues.
Let $\widetilde{\mC}\in\R^{N_t\times N_p}$ respectively $\widetilde{\mP}\in\R^{N_x\times N_p}$ be the matrices that contain only the corresponding eigenvectors as columns.
Then, by inserting \eqref{eq:modes_mos} and \eqref{eq:prefactors} into \eqref{eq:transformation}, the lower-dimensional projection $\widetilde{\mX}$ of the data is given by
\begin{align}\label{eq:projection}
	\widetilde{\mX} = \mX\mT\widetilde{\mC}\widetilde{\mC}^\dagger =:\mX\mat{P}.
\end{align}
In other words, the lower-dimensional projection is calculated by right-multiplying the data set matrix with the projection matrix $\mat{P}\in\R^{N_t\times N_t}$ (which satisfies $\mat{P}^2=\mat{P}$).

To summarise, for a transformation into the POD basis and lower-dimensional projections, no expensive scalar products between modes $\vec{\phi}^i$ and snapshots $\vec{q}^j$ have to be calculated -- all information is already given by the covariance matrix and its eigenvalue problem.

% \begin{itemize}
% 	\item The prefactors $\mXi$ determine the decomposition of the data set in the POD basis modes:
% 		\begin{align}\label{eq:prefactors}
% 			\mX = \mP\mXi\quad\text{with}\quad\mXi=\mP^\dagger\mS\mX=\mL^{\nicefrac12}\mC^\dagger
% 		\end{align}
% 	In the last step of \eqref{eq:prefactors}, the relations \eqref{eq:modes_mos},\eqref{eq:evp_mos} and \eqref{eq:orthonorm_C} have been used.
% 	The $i$-th row of $\mXi$ gives the time series of mode $\vec{\phi}^i$; the $j$-th column is the partition of snapshot $\vec{q}^j$.
% 	\item Lower-dimensional projection is obtained by considering only a number $N_p$ of POD modes (e.g., corresponding to the $N_p$ biggest eigenvalues). Let $\widetilde{\mC}\in\R^{N_t\times N_t}$ be the matrix that contains only the corresponding eigenvectors as columns. Then the lower-dimensional projection $\widetilde{\mX}$ of the data set is given by
% 	\begin{align}
% 		\widetilde{\mX} = \mX\mT\widetilde{\mC}\widetilde{\mC}^\dagger =:\mX\mat{P}
% 	\end{align}
% 	with the projection matrix $\mat{P}\in\R^{N_t\times N_t}$ (with $\mat{P}^2=\mat{P}$).
% 	\item Take-home message: For transformation into POD basis and projections, no scalar products between modes $\vec{\phi}^i$ and snapshots $\vec{q}^j$ have to be calculated -- all information is already contained in $\mL$ and $\mC$.
% \end{itemize}

\subsection{POD Analysis that Maximises the Heat Transport}\label{sec:new_approach}
Up to now, we presented the standard POD approach.
The norm of a snapshot is its generalised energy, $(\vec{q},\vec{q})=\int_\Omega\!\mathrm{d}^2x\,(T^2+u_z^2)$, and the eigenvalues also represent this energy.
But the generalised energy is not a physically meaningful quantity, as it is the sum of temperatures and velocities (this does not change when using non-dimensionalised quantities).
Instead, what \emph{is} a physical quantity that is highly relevant for the Rayleigh--Bénard system is the heat that the fluid transports by convection.
The convective heat transport is measured by the convective Nusselt number
\begin{align}\label{eq:Nuc}
	\Nuc=\Nu-1=\frac{1}{V}\int_\Omega\!\mathrm{d}^2x\,T\,u_z,
\end{align}
assuming that $\langle T\rangle_\Omega = 0$.
This expression suggests that $\Nuc$ can be obtained from a bilinear form
\begin{align}\label{eq:Nuc_bilinearform}
	(\vec{q}^i,\vec{q}^j) = \frac{1}{2V}\int_\Omega\!\mathrm{d}^2x\,(T^iu_z^j + T^ju_z^i)
\end{align}
that generates cross-terms, i.e. products of temperature and velocity.
In the discrete case this corresponds to the matrix
\begin{align}\label{eq:mS}
	\mS = \frac{1}{2V}\left(\begin{matrix}\zero & \mat{W} \\ \mat{W} & \zero \end{matrix} \right).
\end{align}
By utilising this $\Nuc$-inducing matrix in the formalism of the prior sections, it becomes possible for the POD modes and eigenvalues to describe the convective Nusselt number instead of the generalised energy.

The matrix $\mS$ is still Hermitian, which is the most important property utilised in the derivation.
In contrast to \eqref{eq:mS_E}, $\mS$ is not positive definite, though, and thus it does not define a proper scalar product but only a Hermitian bilinear form.
Likewise, $\mS$ does not induce a proper norm but only a metric via $(\vec{q}^i,\vec{q}^i)=\Nuc(\vec{q}^i)$ that may become negative for a snapshot that has a negative convective heat transport (physically speaking, when e.g. strong hot plumes are swept down by the large-scale current).
Correspondingly, the eigenvalues in $\mL$ can become negative, and its inverse square root $\mL^{-\nicefrac12}$ would produce complex POD modes, cf. \eqref{eq:modes_mos}.
Therefore, we introduce the matrix $\mLabs$ that contains the absolutes of the eigenvalues, and utilising it in \eqref{eq:modes_mos} results in the real POD modes
\begin{equation}\label{eq:Nu_modes_mos}
	\mP=\mX\mT\mC\mLabs^{-\nicefrac12}.
\end{equation}
These modes are not normalised to $1$ but to $\pm1$, depending on the signs of the eigenvalues, i.e.
\begin{equation}
	\mP^\dagger\mS\mP = \mLs := \diag\bigl(\sgn{\lambda_1},\dots,\sgn{\lambda_{N_t}}\bigr).
\end{equation}
As a consequence, the amplitudes of the POD modes read
\begin{equation}
	\mXi = \mLabs^{\nicefrac12}\mC^\dagger.
\end{equation}
The lower-dimensional projection \eqref{eq:projection} does not need to be modified.

To conclude, with the Nusselt-inducing $\mS$ \eqref{eq:mS} and trivial modifications like using the absolute eigenvalues, it becomes possible for the POD technique to also describe the convective heat transport.
As the eigenvalues $\lambda_i$ that give the convective heat transport contained in the POD modes $\lambda_i$ may become negative, the modes may be grouped into positive and negative heat transport.
Projections onto e.g. only the positive modes become possible that allow to study the structures that account for the most convective heat transport.

We want to remark that we will not take symmetries into account in our analysis.
Although symmetries may be used to enlarge the data ensemble and there are ways to do so with little additional cost \citep{sirovich90pfa}, in the following sections we will deal mainly with flows that have a strong large-scale circulation where the symmetries are broken.
As a result, when utilising symmetries the POD modes would show no signatures of the large-scale circulations.

\section{Results}\label{sec:results}
We will now apply the theory outlined in the previous section to data sets obtained from direct numerical simulations of two-dimensional convection (section \ref{sec:results_2d}), three-dimensional convection in a cylindrical vessel (section \ref{sec:results_3d}) and to two-dimensional convection with varying Rayleigh number (section \ref{sec:results_rascan}) and with varying aspect ratio (section \ref{sec:results_gammascan}).

\subsection{Two-Dimensional convection}\label{sec:results_2d}
\subsubsection{Data Set}
\begin{figure}
	\centerline{\includegraphics{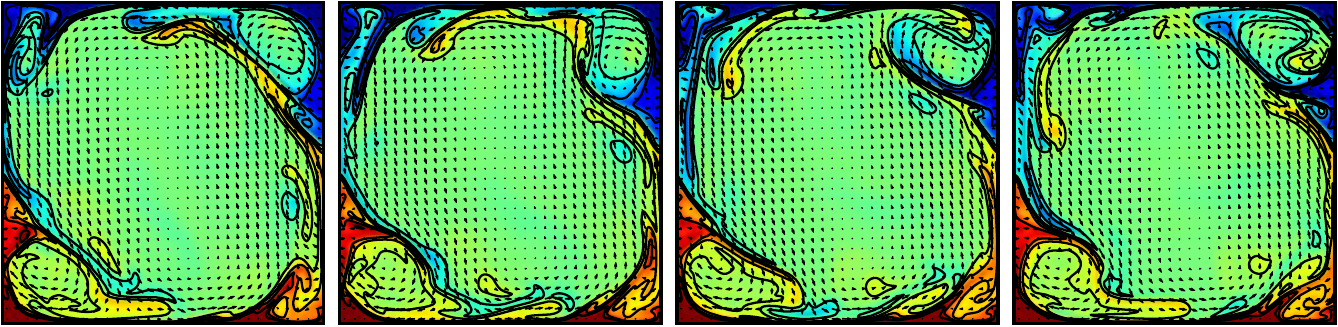}}
	\caption{Data set of two-dimensional Rayleigh--Bénard convection with $\Ra=3\times10^8$, $\Pr=1$ and $\Gamma=1$; the temperature and velocity fields of four consecutive snapshots are shown, separated by $0.5$ free-fall times. The temperature scale and the eight equidistant isothermals have been rescaled to cover $50\,\%$ of the full range, i.e. $-0.25<T<0.25$.\label{fig:rb_dataset}}
\end{figure}
We start with two-dimensional Rayleigh--Bénard convection with fixed side walls and an aspect ratio $\Gamma=1$.
The Rayleigh number is $\Ra=3\times10^8$ and the Prandtl number (kinematic over thermal diffusivity) is $\Pr=1$.
The bottom and top plates are kept at a fixed temperature of $T=0.5$ respectively $T=-0.5$, and the side walls are insulating.
All surfaces are no-slip.
The data ensemble consists of $N_t=500$ snapshots, separated by $0.5$ free-fall times, and each snapshot is resolved with $n_x=736\times736$ grid points.
The numerics are an equidistant pseudo-spectral code where the boundary conditions are enforced by volume penalisation \citep{angot99num,schneider05caf,keetels07jcp}; details can be found in \citet{luelff11njp}.
Four consecutive snapshots of the temperature and velocity fields are shown in figure~\ref{fig:rb_dataset}.
% \begin{itemize}
% 	\item 2D RBC, $\Ra=1.46\times10^8$, $\Pr=1$, $\Gamma=1$, $N_t=500$ snapshots, $736\times736$ equidistant gridpoints ($N_x=1083392$), cf. figure~\ref{fig:dataset}. The numerics are a equidistant pseudospectral code where the boundary conditions are enforced by volume penalization \citep{luelff11njp,angot99num,schneider05caf,keetels07jcp}.
% \end{itemize}

\subsubsection{Eigenvalue Spectrum}
\begin{figure}
	\centerline{\includegraphics{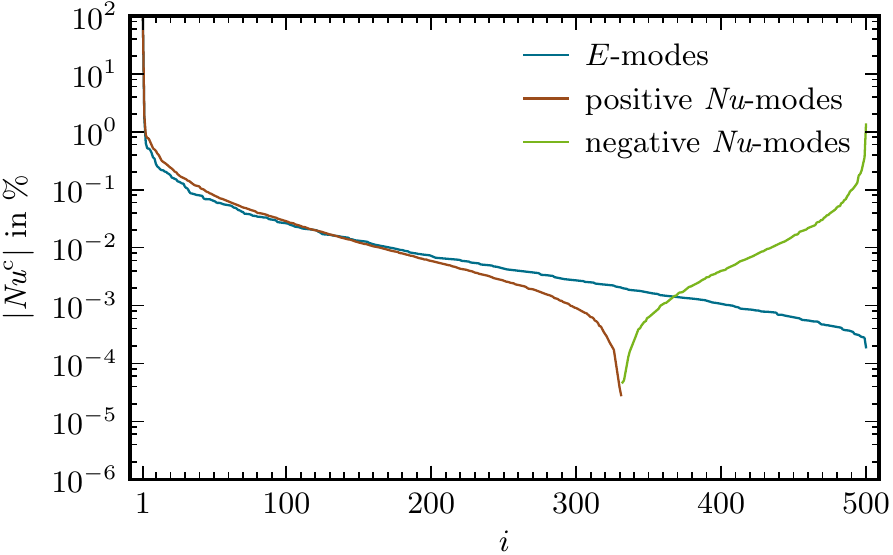}}
	\caption{Convective heat transport spectrum with logarithmic vertical axis for $\Ra=3\times10^8$, $\Pr=1$ and $\Gamma=1$.\label{fig:Lambda_log}}
\end{figure}
In the following sections, we will compare results for the standard energy modes and the proposed Nusselt modes; to distinguish, we will call them $E$- and $\Nu$-modes and add indices to the quantities when needed.
Figure~\ref{fig:Lambda_log} shows the convective heat transport spectrum of the $E$- and $\Nu$-modes, sorted in descending order.
The values are given in percent of the total convective heat transport, which is $\Nuc=34.96$.
The heat transport contained in the $i$-th $\Nu$-mode is directly given as the $i$-th eigenvalue.
The heat transport of the $E$-modes is obtained by projecting the data set onto the $i$-th mode and calculating the heat transport of the projection; it follows that the heat transport spectrum of the $E$-modes is given by the diagonal entries of the matrix $\mC_E^\dagger\mT^\dagger\mX^\dagger\mS_{\Nu}\mX\mT\mC_E$.

\begin{table}
\begin{center}
\begin{tabular}{lrrrrrr}
	  &  $\Nuc(\vec{\phi}^1)\text{ in }\%$  &  $\Nuc(\vec{\phi}^2)\text{ in }\%$  &  $\Nuc(\vec{\phi}^3)\text{ in }\%$  &  $\Nuc(\vec{\phi}^4)\text{ in }\%$  &  $\Nuc(\vec{\phi}^5)\text{ in }\%$  \\[8pt]
	$E$-modes  &  $85.484$  &  $1.746$  &  $0.638$  &  $0.516$  &  $0.514$  \\ 
	$\Nu$-modes  &  $86.840$  &  $1.908$  &  $0.877$  &  $0.773$  &  $0.767$
\end{tabular}
\end{center}
\caption{Convective heat transport of the first $E$- respectively $\Nu$-modes given in percent for $\Ra=3\times10^8$, $\Pr=1$, $\Gamma=1$. The mean heat transport of the data set it $\Nuc=34.964$.\label{tab:ht_modes}}
\end{table}

For the $\Nu$-modes, we obtain $331$ modes with positive heat transport and $169$ negative ones.
The first mode has the highest heat transport and it contains $86.8\,\%$ of the heat transport, and the drop to the second mode is almost $2$ orders of magnitude (cf. also the values given in table~\ref{tab:ht_modes}).
The heat transport of all $E$-modes is positive, and for approximately the first $100$ modes it is below the $\Nu$-modes; after that, the $\Nu$-modes become smaller as they turn negative (note that the absolute value of $\Nuc$ is plotted).
The higher $E$-modes show an exponential decay, indicating their decreasing importance, as is also observed by \cite{bailoncuba10jfm}.

\begin{figure}
	\centerline{\includegraphics{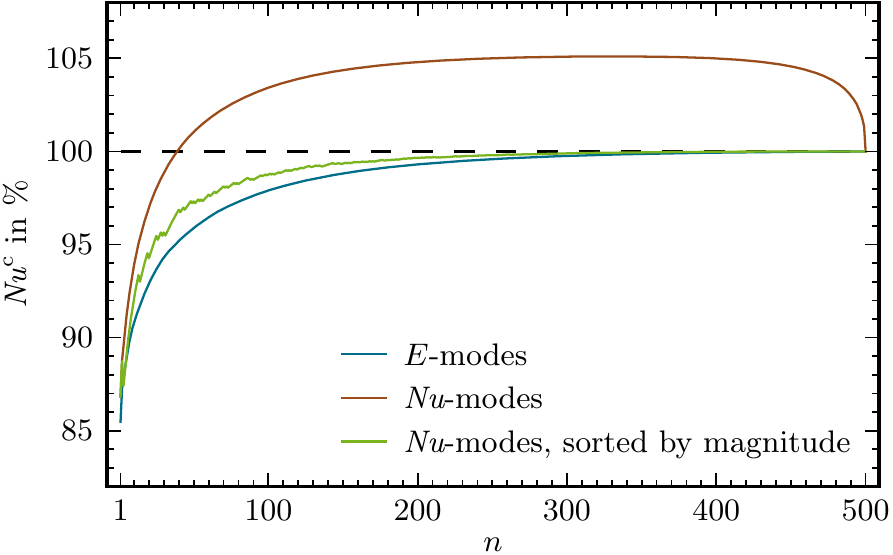}}
	\caption{Integrated convective heat transport spectrum for $\Ra=3\times10^8$, $\Pr=1$ and $\Gamma=1$.\label{fig:Lambda_sum_magn}}
\end{figure}

In figure~\ref{fig:Lambda_sum_magn} the integrated heat transport spectrum is shown in percent, i.e. the $\Nuc$ that is contained in a lower-dimensional projection onto the first $n$ $E$- respectively $\Nu$-modes.
This plot allows to read of the number of modes needed to achieve a projection with a certain amount of heat transport; also cf. table~\ref{tab:nr_modes}.
For the $E$-modes, the convective heat transport starts at $85.5\,\%$ and quickly rises with increasing mode count until in the end it monotonically  converges towards the $100\,\%$ contained in the full data set.
For the $\Nu$-modes, however, the integrated spectrum is not monotonic as there are positive and negative eigenvalues.
Thus, when only considering the positive modes (cf. the brown curve), the projection can actually rise above $100\,\%$ because the negative modes are missing.
The maximum of $105.1\,\%$ is reached when utilising only the $331$ positive modes; after that, the negative modes lower $\Nuc$ again to $100\,\%$.
Likewise, as few as $39$ $\Nu$-modes contain the same convective heat transport as the full data set.

As pointed out earlier, the modes that are used for a lower-dimensional projection can be chosen arbitrarily.
Up to now, we used the $n$ largest ones (in terms of eigenvalue) to show that the same mean heat transport can be achieved with a small number of modes.
One could argue, though, that the negative modes are also important for the physics of the system, and sort the modes by decreasing magnitude of the heat transport.
The result is the green line of figure~\ref{fig:Lambda_sum_magn}.
It is jagged as it is the sum of alternating  positive and negative terms, and now converges towards $100\,\%$ from below. 
But most importantly, the heat transport is always above the one of the $E$-modes, indicating that regardless of the ordering of the modes, the $\Nu$-modes outperform the $E$-modes in terms of describing the heat transport.
For the remainder of the paper, we will only consider sorting the modes by descending value of $\Nuc$.
\begin{table}
\begin{center}
\begin{tabular}{lccccc}
	  &  $n(90\,\%\;\Nuc)$  &  $n(95\,\%\;\Nuc)$  &  $n(99\,\%\;\Nuc)$  &  $n(100\,\%\;\Nuc)$  &  $n(\text{max. }\Nuc)$  \\[8pt]
	$E$-modes  &  $8$  &  $38$  &  $166$  &  $(500)$  &  $(500)$  \\ 
	$\Nu$-modes  &  $4$  &  $13$  &  $32$  &  $39$  &  $331$
\end{tabular}
\end{center}
\caption{Number of modes needed to obtain a specified percentage of convective heat transport for $\Ra=3\times10^8$, $\Pr=1$ and $\Gamma=1$, distinguished into $E$- and $\Nu$-modes. For the $E$-modes, $100\,\%\;\Nuc$ corresponds to the maximum which is obtained for $500$ modes, i.e. when there is no projection.\label{tab:nr_modes}}
\end{table}

\subsubsection{POD Modes}\label{sec:results_2d_modes}
\begin{figure}
	\centerline{\includegraphics{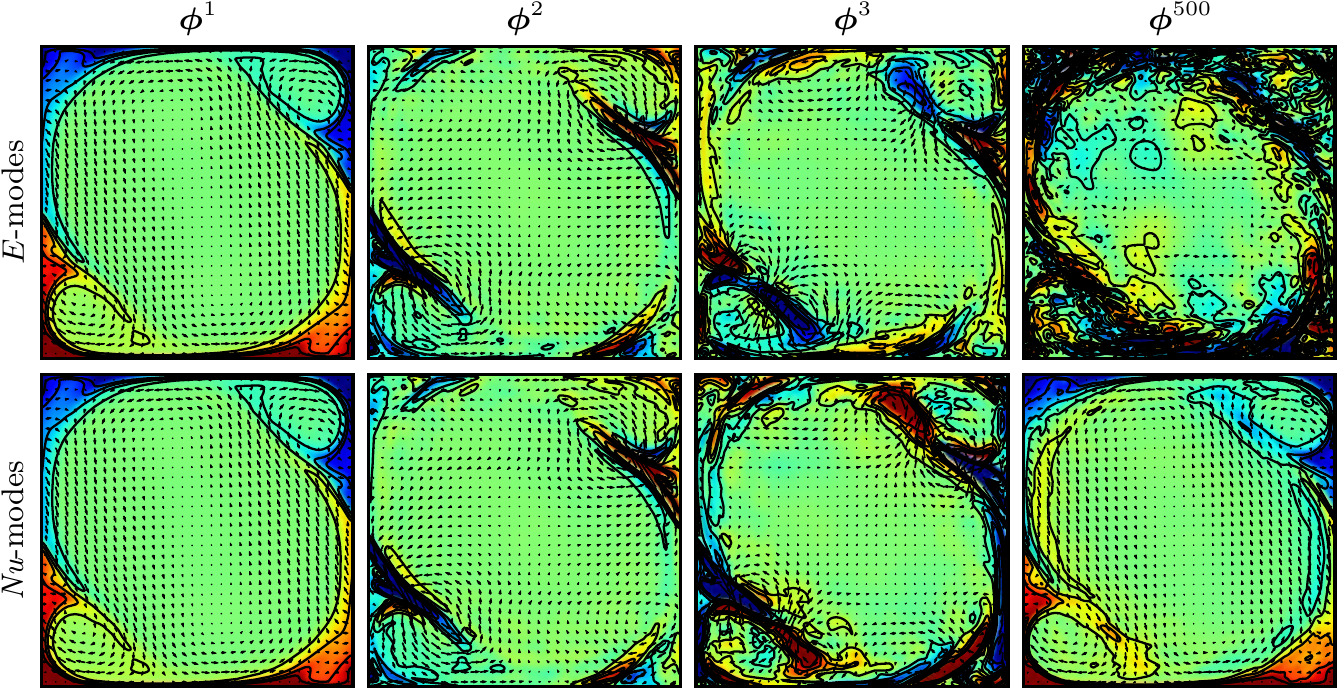}}
	\caption{First three and the last POD mode for $\Ra=3\times10^8$, $\Pr=1$ and $\Gamma=1$. Upper row: $E$-modes; lower row: $\Nu$-modes. As the modes are normalised to $\bigl(\vec{\phi},\vec{\phi}\bigr)=\pm1$, the absolute temperature values become meaningless, and in the visualisations the temperature scale and isothermals are rescaled to cover $50\,\%$ of the total temperature range in all cases (likewise for all POD modes from now on).\label{fig:modes}}
\end{figure}
Figure~\ref{fig:modes} shows some exemplary $E$- and $\Nu$-modes, i.e. the first three and the last one.
In both cases the first mode corresponds to the averaged large-scale circulation conceivable in figure~\ref{fig:rb_dataset}, while the second mode is concentrated near the corner flows and is responsible for their oscillation triggered by passing plumes.
The third mode, however, shows subtle differences between the $E$- and $\Nu$-case, especially near the corner flows: The $E$-mode only favours large values of the fields, and there is cold fluid moving up and moving down, while the $\Nu$-mode also shows more positive correlation between temperature and vertical velocity, i.e. events of positive heat transport.
Also, $\vec{\phi}^3_\Nu$ shows elongated structures near the side wall that correspond to plumes transported by the large-scale current; these structures are not found in $\vec{\phi}^3_E$.
This exemplifies that in contrast to the $E$-modes, the $\Nu$-modes are sensitive to structures of high convective heat transport.
The last $E$-mode is generally featureless and noise-like which again shows that the higher modes become less important.
However, the last $\Nu$-mode has the biggest negative heat transport and shows features similar to the large-scale circulation of $\vec{\phi}^1$, although, upon closer inspection, more regions with negative heat transport are found (e.g., the parts of the corner rolls extending towards the plate and the hot respectively cold vertical structures near the side walls).

% \begin{itemize}
% 	\item Fig. \ref{fig:modes} shows first four $E$- and $\Nu$-modes. Temperature part rescaled to arbitrary values because modes are normalized to $(\vec{\phi},\vec{\phi})=\pm1$, and therefore absolute temperature values become meaningless.
% 	\item First mode identical; for the higher modes, the $\Nu$-modes display structures that represent vertically aligned structures of strong temperatures that correspond to extreme events of heat transport.
% \end{itemize}

\subsubsection{Time Series of Convective Heat Transport}
We will now analyse the performance of the $E$- and $\Nu$-modes in terms of describing the time series of the convective heat transport.
The time series of the full data set corresponds to the diagonal entries of $\mX^\dagger\mS_\Nu\mX$, and from this matrix the time series of lower-dimensional projections are obtained as $\mat{P}^\dagger\mX^\dagger\mS_\Nu\mX\mat{P}$, cf.~\eqref{eq:projection}, where $\mat{P}$ is either $\mat{P}_E$ or $\mat{P}_\Nu$.

\begin{figure}
	\centerline{\includegraphics{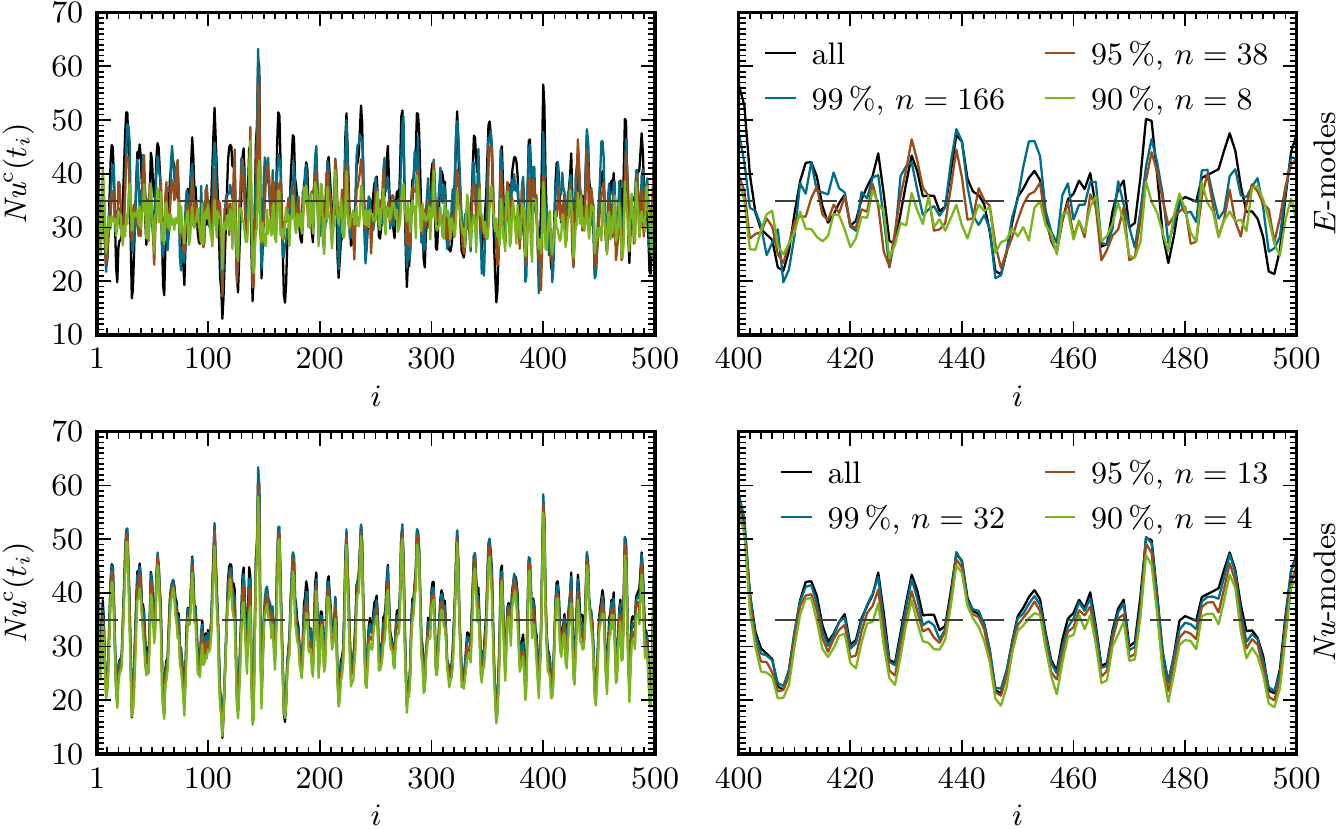}}
	\caption{Time series of the convective heat transport for $\Ra=3\times10^8$, $\Pr=1$ and $\Gamma=1$. Upper row: Time series of $\Nuc$ and of lower-dimensional projections using $\{8,38,166\}$ $E$-modes, corresponding to $\{90,95,99\}\,\%$ heat transport. Upper right panel shows a zoom into the last $100$ time steps. Lower row: Time series of $\Nuc$ and of lower-dimensional projections using $\{4,13,32\}$ $\Nu$-modes, corresponding to $\{90,95,99\}\,\%$ heat transport. Lower right panel again shows zoom.\label{fig:timeseries_Nuc}}
\end{figure}

In figure~\ref{fig:timeseries_Nuc}, the time series of $\Nuc$ is shown, i.e. the convective heat transport contained in each of the $500$ snapshots.
The lower-dimensional projections are chosen such that $\{90,95,99\}\,\%$ of the full convective heat transport are achieved, which implies that a different number of $E$- respectively $\Nu$-modes is used (as given in the caption of the figure).

The $E$-modes do not describe the full signal very well.
Although it is generally seen that with increasing mode count the time series of the projections come closer to the full signal, the deviation of the projection and the full time series is not systematic:
There are snapshots (best seen in the magnification, i.e. upper right panel of figure~\ref{fig:timeseries_Nuc}) where e.g. the $90\,\%$-projection is below and where it is above the true signal, where it is below, above or in between all the other projections, and so on.
Even when utilising $166$ modes (i.e., $33\,\%$ of the available modes) corresponding to $99\,\%$ $\Nuc$, the reconstruction still shows large variations from the true signal.

On the other hand, the lower-dimensional projection utilising the $\Nu$-modes reconstructs the true time series significantly better.
Already the $90\,\%$-projection with as few as $4$ modes describes the full data very well, and the reconstructions become better with increasing mode count.
Also, the deviation is systematic, and in fact it can be shown that \emph{for every time step} the heat transport captured by lower-dimensional projections onto $\Nu$-modes is monotonically increasing with the number of positive modes used.
Furthermore, for a given percentage of $\Nuc$, less $\Nu$-modes than $E$-modes are needed, as also seen in table~\ref{tab:nr_modes}.
To summarise, compared to the standard approach, the proposed $\Nu$-modes show a better performance with fewer modes when describing the time series of the convective heat transport.

However, we want to point out here that it comes as no real surprise that the $\Nu$-modes perform better in describing the heat transport when compared to the $E$-modes, because the $\Nu$-modes were specifically tailored to do so.
Conversely, the $E$-modes would outperform the $\Nu$-modes in describing the generalised energy.
But still, the argument holds that the convective heat transport is a physically meaningful quantity, while the generalised energy is not.

% \begin{itemize}
% 	\item Fig. \ref{fig:timeseries_Nuc} shows time series of convective heat transport
% 	\item $\Nuc$-modes show better performance with fewer modes when compared to the $E$-modes
% \end{itemize}

\subsubsection{Statistics of Local Convective Heat Transport}
\begin{figure}
	\centerline{\includegraphics{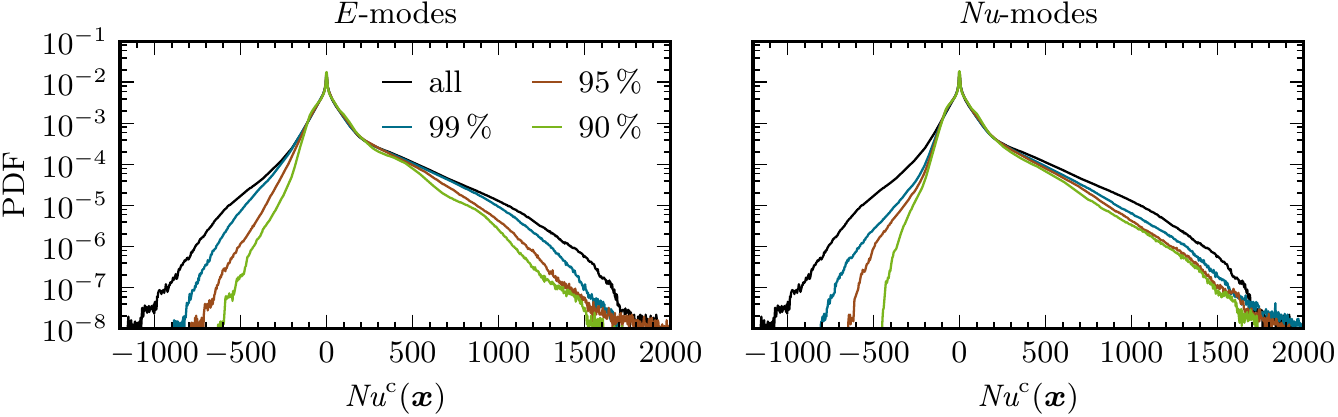}}
	\caption{PDFs of the local convective heat transport $\Nuc(\vec{x})$ for $\Ra=3\times10^8$, $\Pr=1$ and $\Gamma=1$. Left: Full data and projections onto $\{8,38,166\}$ $E$-modes, corresponding to $\{90,95,99\}\,\%$ heat transport. Right: Full data and projections onto $\{4,13,32\}$ $\Nu$-modes, corresponding to $\{90,95,99\}\,\%$ heat transport.\label{fig:pdfs_Nuc}}
\end{figure}

% \begin{itemize}
% 	\item The field of the local convective heat transport
% 		\begin{align}
% 			\Nuc(\vec{x}) = u_z(\vec{x})\theta(\vec{x}) = u_z(\vec{x})\bigl(T(\vec{x}) - \langle T(\vec{x})\rangle_{A,t}\bigr).
% 		\end{align}
% 	gives information about the structures that transport heat, e.g. plumes.
% 	\item Fig. \ref{fig:pdfs_Nuc} shows PDFs of the $\Nuc(\vec{x})$-field. While the $E$-modes cut events in the positive and negative tails, the $\Nu$-modes strongly cut the negative tails and almost completely retain the positive tails.
% \end{itemize}
In order to understand the structures that have the most influence on the heat transport and how they are represented by the POD modes, we will now investigate the local convective heat transport.
It is defined as the field
\begin{align}
	\Nuc(\vec{x}) = u_z(\vec{x})\,\theta(\vec{x}) = u_z(\vec{x})\bigl(T(\vec{x}) - \langle T(\vec{x})\rangle_{A,t}\bigr),
\end{align}
where $\langle\cdot\rangle_{A,t}$ is an average over horizontal planes and time and $\theta(\vec{x})$ is the temperature deviation from the mean which is the quantity relevant for buoyancy.
Figure~\ref{fig:pdfs_Nuc} shows the probability density function (PDF) of $\Nuc(\vec{x})$.
The black curve of the full data set indicates that the PDF is skewed towards positive values, which is reasonable since positive correlation of temperature and vertical velocity (i.e., rising hot fluid or falling cold fluid) is the generic behaviour of Rayleigh--Bénard convection.
On the other hand, negative events happen less often, e.g. hot plumes that are coincidentally swept downwards by the large-scale current.
The modal value $\Nuc(\vec{x})=0$ of the PDF corresponds to the no-slip boundary conditions where $u_z=0$.

The PDFs of the lower-dimensional projections instructively show the difference between the $E$- and $\Nu$-modes:
For the $E$-modes, events in both tails of the PDF are cut, with steeper tails for lesser number of modes.
The PDFs of the projections that use the $\Nu$-modes almost completely retain the positive tails, while the negative tails fall of more steeply.
Thus, the $\Nu$-projection removes the strong negative events from the data set while still retaining almost all positive extreme events, and thus fewer modes suffice to achieve the same heat transport compared to the $E$-modes.
We see that in contrast to the usual ansatz, the $\Nu$-modes are sensitive to the heat transport, which we utilised by only considering the most positive modes in the projections.

\subsection{Three-Dimensional Cylindrical Convection}\label{sec:results_3d}

We now turn to three-dimensional Rayleigh--Bénard convection in a cylindrical vessel.
As the theory in section \ref{sec:theory} was developed with respect to two-dimensional convection, we will first outline the needed adaptations, and then show the eigenvalue spectrum and Nusselt modes.

The cylindrical convection is described in azimuthal, radial and axial coordinates $(\varphi,r,z)$.
Due to incompressibility, two velocity components are enough to fully describe the fluid state, e.g. $u_z$ and $u_{\varphi}$, and the state vector becomes
\begin{align}
	\vec{q}=\bigl(T_1,\dots,T_{n_x},u_{\varphi,1},\dots,u_{\varphi,n_x},u_{z,1},\dots,u_{z,n_x}\bigr)^T\in\R^{N_x}
\end{align}
with $N_x=3\,n_x=3\,n_\varphi\times n_r\times n_z$ spatial degrees of freedom.
The Nusselt number inducing bilinear form then becomes
\begin{align}
	\mS = \frac{1}{2V}\left(\begin{matrix}\zero & \zero & \mat{W} \\ \zero & \zero & \zero \\ \mat{W} & \zero & \zero\end{matrix}\right)\in\R^{N_x\times N_x}
\end{align}
with $\mat{W}$ containing the spatial weights as in \eqref{eq:mS_E}.
While this bilinear form is still Hermitian, it is not non-degenerate any more because it does not depend on information about the horizontal velocities, i.e. $\nicefrac13$ of the data is not accounted for.
The proposed $\Nu$-method still works and it produces $N_t$ meaningful orthonormal modes from $N_t$ snapshots, as long as there are enough spatial degrees of freedom, i.e. $\frac{2}{3}N_x>N_t$ instead of the usual $N_x>N_t$.
This can be shown by using estimations on the rank of a product of matrices, i.e. the covariance matrix \eqref{eq:evp_mos}.
For the data sets we are considering, the requirement $\frac{2}{3}N_x>N_t$ is usually fulfilled.
Lower-dimensional projections and the POD modes of course contain the horizontal velocities, as both are constructed as linear combinations of the snapshots (cf. section \ref{sec:mos}, equations \eqref{eq:modes_mos} and \eqref{eq:projection}).

\begin{figure}
	\centerline{\includegraphics[width=1.0\textwidth]{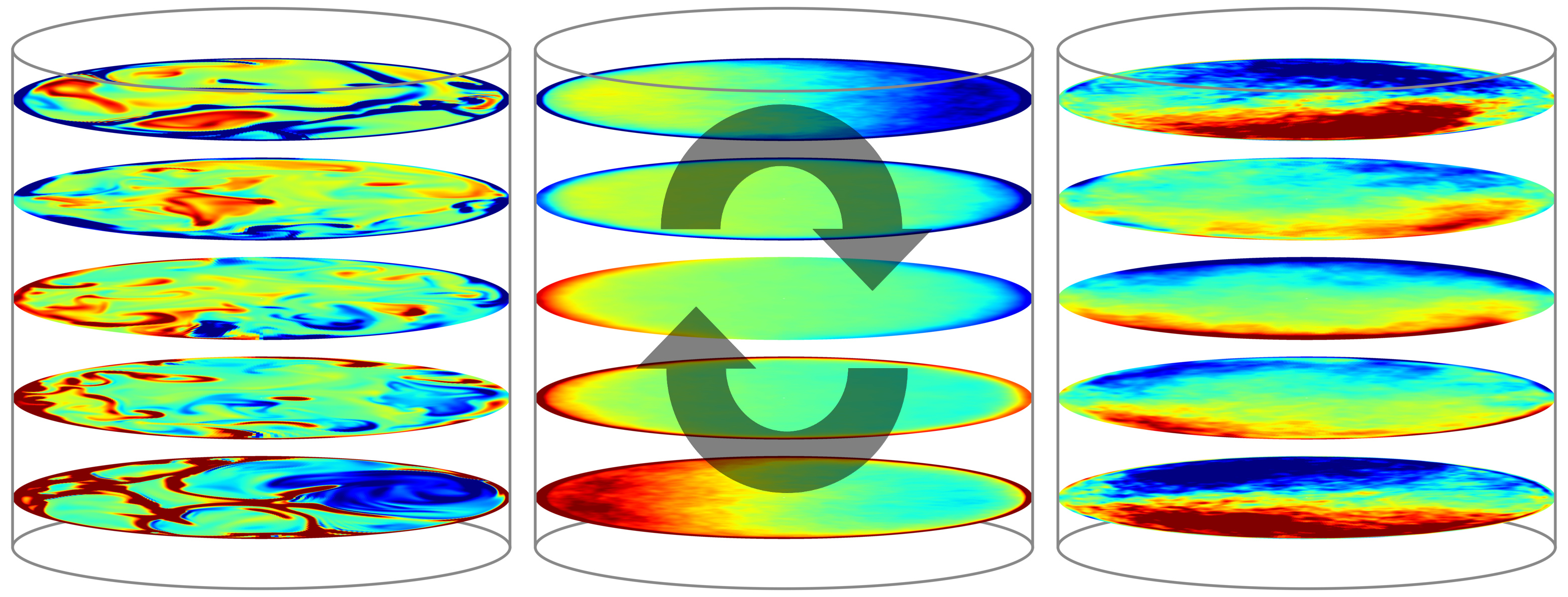}}
	\caption{Temperature field of three-dimensional convection in a cylindrical vessel with $\Ra=2\times10^8$, $\Pr=1$ and $\Gamma=1$ (diameter over height). Horizontal slices at $z\in\{0.1,0.3,0.5,0.7,0.9\}$ are shown. The arrows in the middle panel indicate the large-scale circulation. To enhance the discernible features of the fields, different temperature scales where used; the shown temperature range is given in terms of the temperature standard deviation $\sigma_T$. Left: First snapshot $\vec{q}^1$ with temperature range $1\,\sigma_T$. Middle: First $\Nu$-mode $\vec{\phi}^1_\Nu$ with temperature range $1\,\sigma_T$. Right: Second $\Nu$-mode $\vec{\phi}^2_\Nu$ with temperature range $3\,\sigma_T$.\label{fig:viz_cyl_snap_PODS_arrow}}
\end{figure}

We will now analyse cylindrical convection with $\Ra=2\times10^8$, $\Pr=1$ and aspect ratio (diameter over height) $\Gamma=1$.
The data set consists of $N_t=870$ snapshots, separated by $1$ free-fall time unit, and the resolution of the staggered cylindrical grid is $n_x=384\times192\times384$.
Details of the numerics are given in \cite{verzicco03jfm}.
Figure~\ref{fig:viz_cyl_snap_PODS_arrow} (left) shows a snapshot of the temperature field.

\begin{figure}
	\centerline{\includegraphics{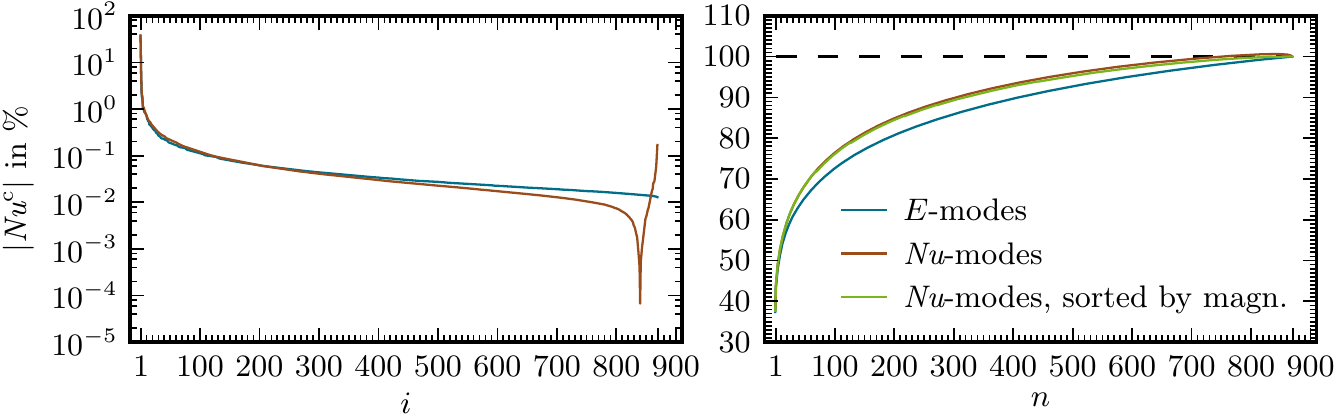}}
	\caption{Heat transport spectrum (left) and integrated heat transport spectrum (right) for cylindrical convection with $\Ra=2\times10^8$, $\Pr=1$ and $\Gamma=1$\label{fig:cyl_spectrum_Nu}.}
\end{figure}

\begin{table}
\begin{center}
\begin{tabular}{lrrrrrr}
	  &  $\Nuc(\vec{\phi}^1)\text{ in }\%$  &  $\Nuc(\vec{\phi}^2)\text{ in }\%$  &  $\Nuc(\vec{\phi}^3)\text{ in }\%$  &  $\Nuc(\vec{\phi}^4)\text{ in }\%$  &  $\Nuc(\vec{\phi}^5)\text{ in }\%$  \\[8pt]
	$E$-modes, diffusive  &  $37.358$  &  $5.862$  &  $1.847$  &  $1.747$  &  $1.015$  \\ 
	$E$-modes, free fall  &  $37.493$  &  $5.880$  &  $1.836$  &  $1.743$  &  $1.037$  \\
	$\Nu$-modes  &  $37.685$  &  $6.140$  &  $2.057$  &  $1.964$  &  $1.134$
\end{tabular}
\end{center}
\caption{Convective heat transport of the first $E$- respectively $\Nu$-modes given in percent for cylindrical convection with $\Ra=2\times10^8$, $\Pr=1$, $\Gamma=1$. The mean heat transport of the data set it $\Nuc=37.653$. The $E$-modes have been further distinguished into diffusive and free fall velocity scales, cf. text.\label{tab:ht_modes_cyl}}
\end{table}

The heat transport spectrum of the $\Nu$- and $E$-modes is displayed in figure~\ref{fig:cyl_spectrum_Nu} (left), with the $E$-modes corresponding to the generalised energy $T^2+u_\varphi^2+u_z^2$.
Again, the first mode contains most of the convective heat transport, but compared to the two-dimensional convection from before, the spectrum falls of less steeply.
Also, the drop from the first to the second mode is only a factor of about $6$ (cf. table~\ref{tab:ht_modes_cyl}), compared to a factor of $46$, which indicates a more dominant large-scale circulation in the two-dimensional case.
For the $\Nu$-modes there are $841$ positive and $29$ negative modes; we relate this to the observation that in three dimensions negative events of convective heat transport are much rarer:
Loosely speaking, due to the stronger confinement in two dimensions strong hot plumes are more often swept downwards which results in more negative events and thus more negative modes.
On the other hand, in three dimensions hot fluid being swept down has another lateral dimension into which it can leave the large-scale circulation.

From the integrated eigenvalue spectrum (figure~\ref{fig:cyl_spectrum_Nu} (right) and table~\ref{tab:nr_modes_cyl}) it becomes clear that the $\Nu$-modes outperform the $E$-modes in describing the heat transport, as for every projection onto $n$ modes, the $\Nu$-modes capture a greater share of the convective heat transport.
The difference is not as big as in the two-dimensional case (figure~\ref{fig:Lambda_sum_magn}), though, because there are only few $\Nu$-modes that negatively contribute to the heat transport.
It is also seen that with few negative modes, the order in which the modes are integrated becomes less important, and both orderings (descending order and descending magnitude) perform equally well.

\begin{table}
\begin{center}
\begin{tabular}{lccccc}
	  &  $n(90\,\%\;\Nuc)$  &  $n(95\,\%\;\Nuc)$  &  $n(99\,\%\;\Nuc)$  &  $n(100\,\%\;\Nuc)$  &  $n(\text{max. }\Nuc)$  \\[8pt]
	$E$-modes, diffusive  &  $411$  &  $593$  &  $802$  &  $(870)$  &  $(870)$  \\ 
	$E$-modes, free fall  &  $396$  &  $578$  &  $796$  &  $(870)$  &  $(870)$  \\ 
	$\Nu$-modes  &  $305$  &  $462$  &  $670$  &  $751$  &  $841$
\end{tabular}
\end{center}
\caption{Number of modes needed to obtain a specified percentage of convective heat transport for cylindrical convection with $\Ra=2\times10^8$, $\Pr=1$ and $\Gamma=1$, distinguished into $E$- and $\Nu$-modes. For the $E$-modes, $100\,\%\;\Nuc$ corresponds to the maximum which is obtained for $870$ modes, i.e. when there is no projection.\label{tab:nr_modes_cyl}}
\end{table}

The first two $\Nu$-mode are shown in the middle and right panel of figure~\ref{fig:viz_cyl_snap_PODS_arrow}.
The first mode represents the large-scale circulation parallel to the drawing plane and shows hot fluid rising on the left side and cold fluid falling down on the right side of the vessel.
The second mode acts perpendicular to the plane of the large-scale circulation, and is stronger near the bottom and top plates than in the middle.
Also, upon closer inspection it seems that it is twisting with the vertical coordinate, as the sample for $z=0.1$ appears rotated counter-clockwise and for $z=0.9$ rotated clockwise. 
We therefore attribute this mode to the so-called twisting and sloshing of the large-scale circulation, as also analysed in e.g. \cite{zhou09jfm}.

\subsubsection{Comparing Different Generalised Energies}
% \begin{itemize}
% 	\item The construction of the $\Nu$-covariance matrix is clear. For the construction of the covariance matrix with the generalized energy, though, there are a number of possibilities how to define a generalized energy:
% 	\begin{itemize}
% 		\item $\vec{u}^2$, i.e. the turbulent kinetic energy as done in \citet{bailoncuba10jfm}.
% 		\item $T^2+u_z^2$, i.e. the same two fields as used in the $\Nu$-analysis
% % 		\item $T^2+\vec{u}^2$, i.e. temperature variance plus kinetic energy
% 		\item Also the kind of non-dimensionalisation is arbitrary, e.g. free-fall vs. diffusive velocity scales
% 	\end{itemize}
% 	\item The integrated spectrum below shows that the heat that the $E$-modes transport is largely uncorrelated with the chosen kind of generalized energy, and is always below the integrated spectrum of the $\Nu$-modes
% \end{itemize}
While the choice of the bilinear form that induces the Nusselt number is clear, there is some freedom in defining a generalised energy.
For the cylindrical case, we have used the generalised energy $T^2+u_\varphi^2+u_z^2$, with the velocities measured in diffusive units. 
As this generalised energy utilises the temperature field and two of the velocity components that are needed to fully identify the velocity field, this corresponds to the generalised energy $T^2+u_z^2$ that was used for the two-dimensional data set in section~\ref{sec:results_2d}.

Other conceivable choices to define a generalised energy are the kinetic energy $\vec{u}^2$, as was done in \cite{bailoncuba10jfm}, the total $L^2$-norm of the fields ($T^2+\vec{u}^2$), or $T^2+u_z^2$, i.e. the same fields that the Nusselt-inducing bilinear form depends upon.
Also, the way of non-dimensionalising the quantities is arbitrary, with possible choices including diffusive and free-fall velocities.

\begin{figure}
	\centerline{\includegraphics{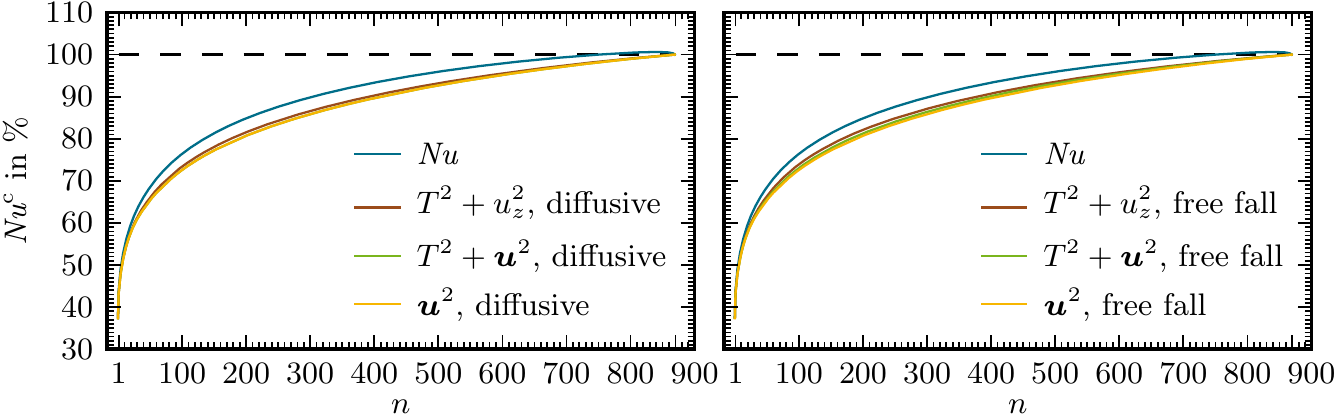}}
	\caption{Integrated heat transport spectrum using different generalised energies for cylindrical convection with $\Ra=2\times10^8$, $\Pr=1$ and $\Gamma=1$.\label{fig:cy_spectrum_diff_E}}
\end{figure}

Figure~\ref{fig:cy_spectrum_diff_E} compares the integrated heat transport spectra of the Nusselt modes and $E$-modes using different generalised energies as well as diffusive and free-fall velocities.
We find that, first, the different generalised energies and different non-dimensionalisations show approximately the same behaviour, and second and most importantly, the heat transport of the Nusselt modes exceeds the $E$-modes in all cases.
On a side note, the curves for $T^2+\vec{u}^2$ and $\vec{u}^2$ in diffusive units lie on top of each other due to the different orders of magnitude of the quantities: While temperature is given relative to the plate temperatures and therefore $T^2=\mathcal{O}(1)$, the velocity of the large-scale circulation is roughly $\vec{u}^2=\mathcal{O}(\Ra)=\mathcal{O}(10^8)$.

\subsection{Two-Dimensional Convection with Varying Rayleigh Number}\label{sec:results_rascan}

\begin{figure}
	\centerline{\includegraphics{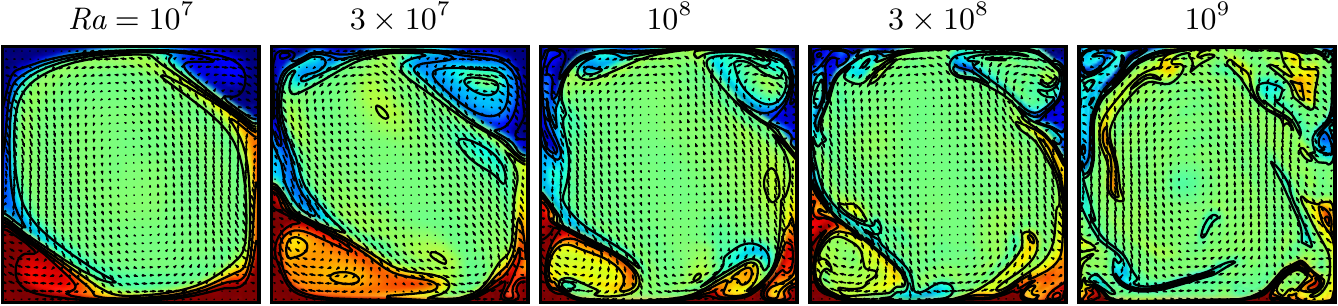}}
	\caption{Snapshots for five different Rayleigh numbers with $\Pr=1$ and $\Gamma=1$. The colour scale is chosen as detailed in the caption of figure~\ref{fig:rb_dataset}.\label{fig:rascan_snapshots}}
\end{figure}

In this section we will investigate the $\Nu$-modes and their heat transport spectrum for two-dimensional convection with varying Rayleigh number $\Ra\in\{10^7,3\times10^7,10^8,3\times10^8,10^9\}$ and with a fixed Prandtl number $\Pr=1$ and aspect ratio $\Gamma=1$.
The boundary conditions are as before, i.e. adiabatic side walls and all surfaces are no-slip.
As can be seen from the snapshots of the flow and temperature fields given in figure~\ref{fig:rascan_snapshots}, varying the Rayleigh number in the range $10^7\le\Ra\le10^9$ has a considerable impact on the structures that appear in the convection cell, and with rising $\Ra$ the small scales become more important, which should also manifest in the POD analysis.
Upon closer inspection of the dynamics, for $\Ra=10^7$ the convection shows regular reversals triggered by the linear growth of the corner rolls that   occur with a distinct frequency \citep{sugiyama10prl}; this is in contrast to the irregularly triggered reversals induced by plumes that are observed for convection with stress-free boundaries \citep{petschel11pre}.
Convection with $\Ra\in\{3\times10^7,10^8,3\times10^8\}$ shows a stable large-scale circulation without reversals, and for $\Ra=10^9$ the flow becomes more irregular and the large-scale structure is less predominant.

\begin{figure}
	\centerline{\includegraphics{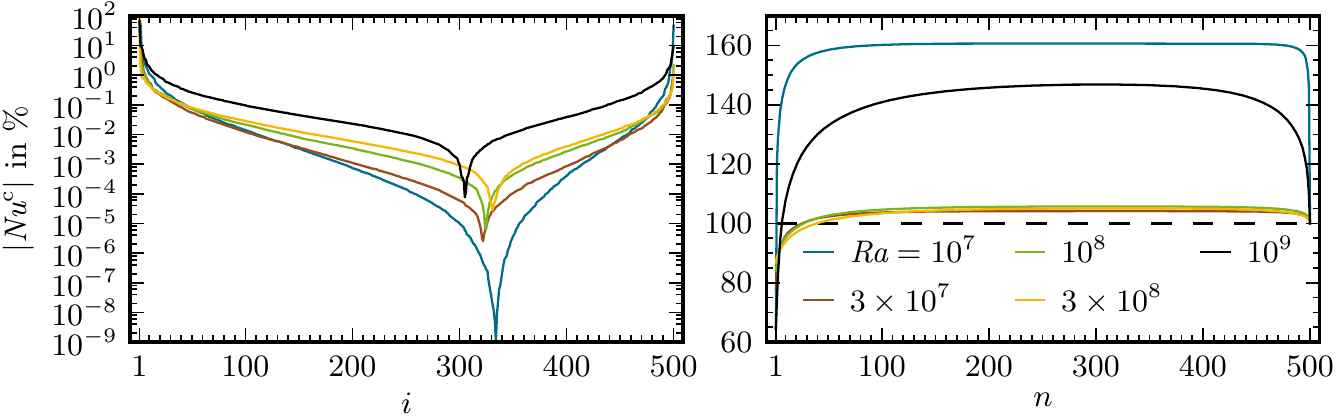}}
	\caption{Eigenvalue spectrum (left) and integrated eigenvalue spectrum (right) of the $\Nu$-modes for different Rayleigh numbers with $\Pr=1$ and $\Gamma=1$.\label{fig:rascan_spectrum}}
\end{figure}

\begin{table}
\begin{center}
\begin{tabular}{lrrrrr}
	  &  $\Ra=10^7$  &  $3\times10^7$  &  $10^8$  &  $3\times10^8$  &  $10^9$  \\[8pt]
	$\Nuc=\sum_j\lambda_j$  &  $10.966$  &  $16.429$  &  $24.825$  &  $34.964$  &  $52.861$  \\[6pt]
	$\lambda_{1}$ in \%  &  $64.340$  &  $77.974$  &  $84.126$  &  $86.840$  &  $64.653$  \\ 
	$\lambda_{2}$ in \%  &  $56.645$  &  $6.718$  &  $2.571$  &  $1.908$  &  $9.617$  \\ 
	$\lambda_{3}$ in \%  &  $8.184$  &  $3.508$  &  $2.297$  &  $0.877$  &  $8.787$  \\ 
	$\lambda_{4}$ in \%  &  $5.183$  &  $2.329$  &  $1.566$  &  $0.773$  &  $6.924$  \\ 
	$\lambda_{5}$ in \%  &  $3.482$  &  $1.307$  &  $1.053$  &  $0.767$  &  $4.464$  \\[6pt] 
	$\lambda_{498}$ in \%  &  $-2.417$  &  $-0.373$  &  $-0.318$  &  $-0.278$  &  $-3.566$  \\ 
	$\lambda_{499}$ in \%  &  $-5.344$  &  $-0.546$  &  $-0.423$  &  $-0.360$  &  $-5.332$  \\ 
	$\lambda_{500}$ in \%  &  $-46.007$  &  $-1.230$  &  $-2.097$  &  $-1.333$  &  $-9.654$ \\[6pt]
	$\sum\limits_{\lambda_j>0}\lambda_j$ in \%  &  $160.673$  &  $104.177$  &  $105.731$  &  $105.105$  &  $146.858$  \\
\end{tabular}
\end{center}
\caption{Heat transport of different $\Nu$-modes for different Rayleigh numbers and $\Pr=1$, $\Gamma=1$. Single eigenvalues are given in percent of the total heat transport, i.e. $100\,\%\times\frac{\lambda_i}{\sum_j\lambda_j}$.\label{tab:rascan_relative_heattransport}}
\end{table}

\begin{figure}
	\centerline{\includegraphics{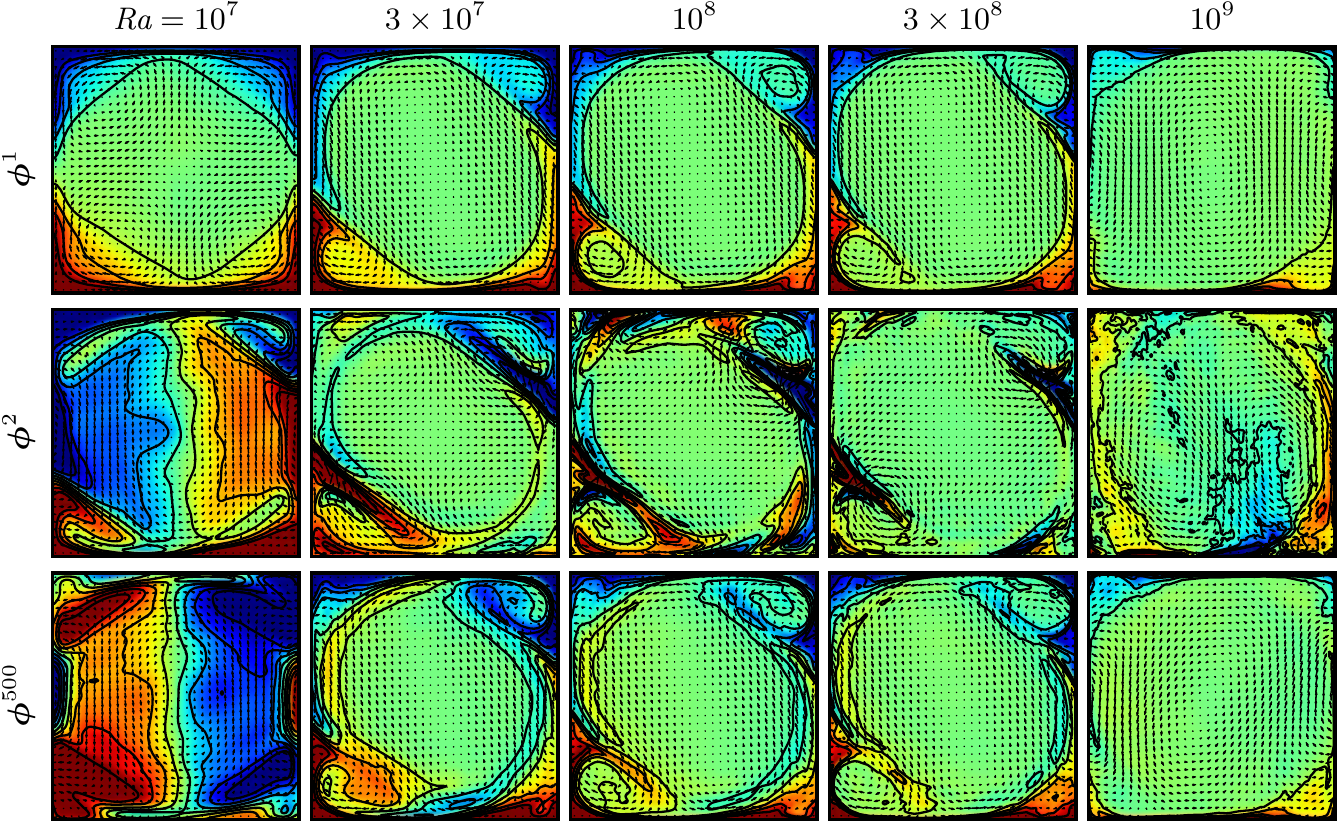}}
	\caption{$\Nu$-modes $\vec{\phi}^1$, $\vec{\phi}^2$ and $\vec{\phi}^{500}$ (i.e., the two most positive and the most negative mode) for five different Rayleigh numbers and $\Pr=1$, $\Gamma=1$.\label{fig:rascan_pods}}
\end{figure}

Figure~\ref{fig:rascan_spectrum} (left) shows the eigenvalue spectra for the different Rayleigh numbers.
The spectra for the higher Rayleigh numbers are flatter, indicating that more modes are needed to describe the fluid flow -- this is in line with the increasing complexity and fine-scale structures suggested by the snapshots.
Interestingly, for $\Ra=10^7$ the first and second eigenvalue are similar (also cf. table~\ref{tab:rascan_relative_heattransport}), while for the other Rayleigh numbers the drop from the first eigenvalue is bigger.
This is related to the reversals observed at $\Ra=10^7$, as the reorientation of the corner flow structure cannot be described by one mode alone, in contrast to the non-reversing large-scale circulations of the other Rayleigh numbers.
For $\Ra=10^9$, the relative share of the heat transport contained in the first mode is rather small and the spectrum is flatter when compared with the other Rayleigh numbers, and also the first mode is less structured, as seen in figure~\ref{fig:rascan_pods}.
This is in line with the observation that for high $\Ra$, the large-scale circulation becomes less important while the small-scale fluctuations become stronger and stronger.

The integrated eigenvalue spectrum in figure~\ref{fig:rascan_spectrum} (right) is for $\Ra\in\{3\times10^7, 10^8, 3\times10^8\}$ similar to the case discussed in section~\ref{sec:results_2d}, and the maximal values also are around $105\,\%$.
For $\Ra=10^7$ and $\Ra=10^9$, however, the integrated spectra take maximal values of $161\,\%$ respectively $147\,\%$, and also $100\,\%\;\Nuc$ are reached with considerably less modes.
We think this is to due the fact that for these Rayleigh numbers, there is no clear large-scale structure, and thus more strong modes are needed to build the erratic fluid behaviour.
Table~\ref{tab:rascan_relative_heattransport} also shows that the positive and negative eigenvalues are much stronger.
Therefore a projection onto only the positive modes results in a drastically increased heat transport because the strong negative modes are neglected.

Figure~\ref{fig:rascan_pods} shows the two most positive and the most negative POD modes.
For $\Ra\in\{3\times10^7,10^8,3\times10^8\}$ the first mode represents the counter-clockwise large-scale circulation.
At $\Ra=10^9$ the convection is dominated by small-scale plumes, and the large-scale current seen in the first POD is less structured; for example, here the corner flows are almost completely missing from $\vec{\phi}^1$.
In contrast, the first mode at $\Ra=10^7$ is horizontally symmetric due to the regularly reversing large-scale current; the reversals seems to be described to a large extent by $\vec{\phi}^1$, $\vec{\phi}^2$ and $\vec{\phi}^{500}$.
The second mode at $\Ra\in\{3\times10^7,10^8,3\times10^8\}$ is again responsible for the deformation and oscillation of the corner rolls, with larger rolls for lower Rayleigh numbers.
At $\Ra=10^9$ the second mode suggests a diagonal two-roll structure, which indicates that the flow is not in a simple one-roll state.
While the last mode of $\Ra=10^7$ has a large eigenvalue and is involved in the reversal of the large-scale current, for the other Rayleigh numbers the last mode is similar to the first one with subtle differences resulting in a negative heat transport, as already discussed in section~\ref{sec:results_2d_modes}.

\subsection{Two-Dimensional Convection with Varying Aspect Ratio}\label{sec:results_gammascan}

\begin{figure}
	\centerline{\includegraphics{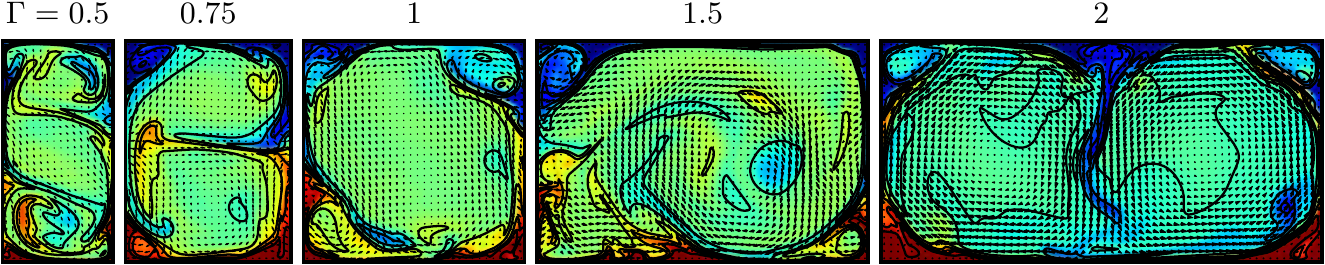}}
	\caption{Snapshots for five different aspect ratios with $\Ra=10^8$ and $\Pr=1$.\label{fig:gammascan_snapshots}}
\end{figure}
In two dimensional convection, the large-scale structure is strongly depending on the geometry, i.e. the aspect ratio $\Gamma$ of the fluid vessel.
We therefore want to investigate the main modes and their heat transport for a varying $\Gamma$.
To this end we performed direct numerical simulations of two-dimensional Rayleigh--Bénard convection with $\Ra=10^8$, $\Pr=1$ and $\Gamma\in\{0.5,0.75,1,1.5,2\}$; snapshots of the flow structure are shown in figure~\ref{fig:gammascan_snapshots}.
For $\Gamma=0.5$, the convection shows an oscillating three-roll structure where the upper and lower roll periodically grow until a reversal is triggered, cf.~\citet{poel11pre}.
In the $\Gamma=0.75$-case, two vertically stacked convection rolls form a stable large-scale circulation, and plumes cannot reach directly from one plate to the other; this results in a drastically reduced $\Nuc$, cf.~table~\ref{tab:gammascan_relative_heattransport}.
For an aspect ratio of $\Gamma=1.5$, the convection does not form a distinct large-scale structure, and instead the convection becomes dominated by erratic small-scale turbulence -- neither a one- nor a two-cell structure is established by this system.
The largest aspect ratio $\Gamma=2$ again shows a stable two-cell structure, and in the middle of the upper plate a cold plume detaches that oscillates horizontally.

\begin{figure}
	\centerline{\includegraphics{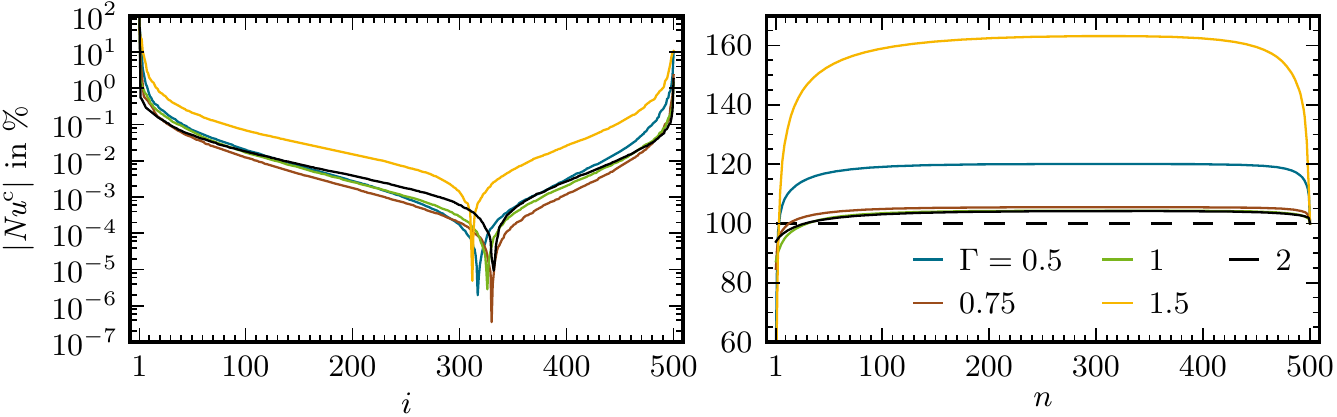}}
	\caption{Eigenvalue spectrum (left) and integrated eigenvalue spectrum (right) for different aspect ratios with $\Ra=10^8$ and $\Pr=1$.\label{fig:gammascan_spectrum}}
\end{figure}
\begin{table}
\begin{center}
\begin{tabular}{lrrrrr}
	  &  $\Gamma=0.5$  &  $0.75$  &  $1.0$  &  $1.5$  &  $2.0$  \\[8pt]
	$\Nuc=\sum_j\lambda_j$  &  $24.437$  &  $20.298$  &  $24.521$  &  $24.787$  &  $24.884$  \\[6pt]
	$\lambda_{1}$ in \%  &  $57.916$  &  $84.738$  &  $87.234$  &  $43.570$  &  $93.850$  \\ 
	$\lambda_{2}$ in \%  &  $26.678$  &  $6.253$  &  $1.686$  &  $24.656$  &  $0.549$  \\ 
	$\lambda_{3}$ in \%  &  $12.422$  &  $3.072$  &  $1.338$  &  $23.467$  &  $0.509$  \\ 
	$\lambda_{4}$ in \%  &  $3.576$  &  $1.871$  &  $1.026$  &  $10.736$  &  $0.446$  \\ 
	$\lambda_{5}$ in \%  &  $2.530$  &  $0.604$  &  $0.927$  &  $8.097$  &  $0.380$  \\[6pt] 
	$\lambda_{498}$ in \%  &  $-1.216$  &  $-0.456$  &  $-0.255$  &  $-7.491$  &  $-0.152$  \\ 
	$\lambda_{499}$ in \%  &  $-2.251$  &  $-0.638$  &  $-0.323$  &  $-9.463$  &  $-0.242$  \\ 
	$\lambda_{500}$ in \%  &  $-9.303$  &  $-2.301$  &  $-1.420$  &  $-10.634$  &  $-1.806$  \\[6pt]
	$\sum_{\lambda_j>0}\lambda_j$ in \%  &  $120.052$  &  $105.485$  &  $104.322$  &  $163.179$  &  $104.178$
\end{tabular}
\end{center}
\caption{Heat transport of different modes for different aspect ratios and $\Ra=10^8$, $\Pr=1$. Single eigenvalues are given in percent of the total heat transport.\label{tab:gammascan_relative_heattransport}}
\end{table}
The eigenvalues (cf. figure~\ref{fig:gammascan_spectrum} and table~\ref{tab:gammascan_relative_heattransport}) clearly show a difference between the cases with a stable pattern ($\Gamma\in\{0.75,1,2\}$ and those with an unstable or oscillating pattern ($\Gamma\in\{0.5,1.5\}$:
The former cases with the strongest large-scale circulation show the biggest drop from the first eigenvalue.
For the latter, the drop in the heat transport is not very steep, indicating that more modes are contained in the reversal process of $\Gamma=0.5$ or the incoherent turbulence of $\Gamma=1.5$.
Also, the most negative mode contains quite a big share of the heat transport, indicating that in these flows events of negative heat transport are much more common.
These observations agree with the cases $\Ra\in\{10^7,10^9\}$ of section~\ref{sec:results_gammascan}.
Furthermore, the $\Gamma=1.5$-case shows an interesting eigenvalue spectrum that lies above all the other aspect ratios, which clearly separates it from the other aspect ratios; this is comparable to the eigenvalue spectrum for $\Ra=10^9$ in figure~\ref{fig:rascan_spectrum} (left).

\begin{figure}
	\centerline{\includegraphics{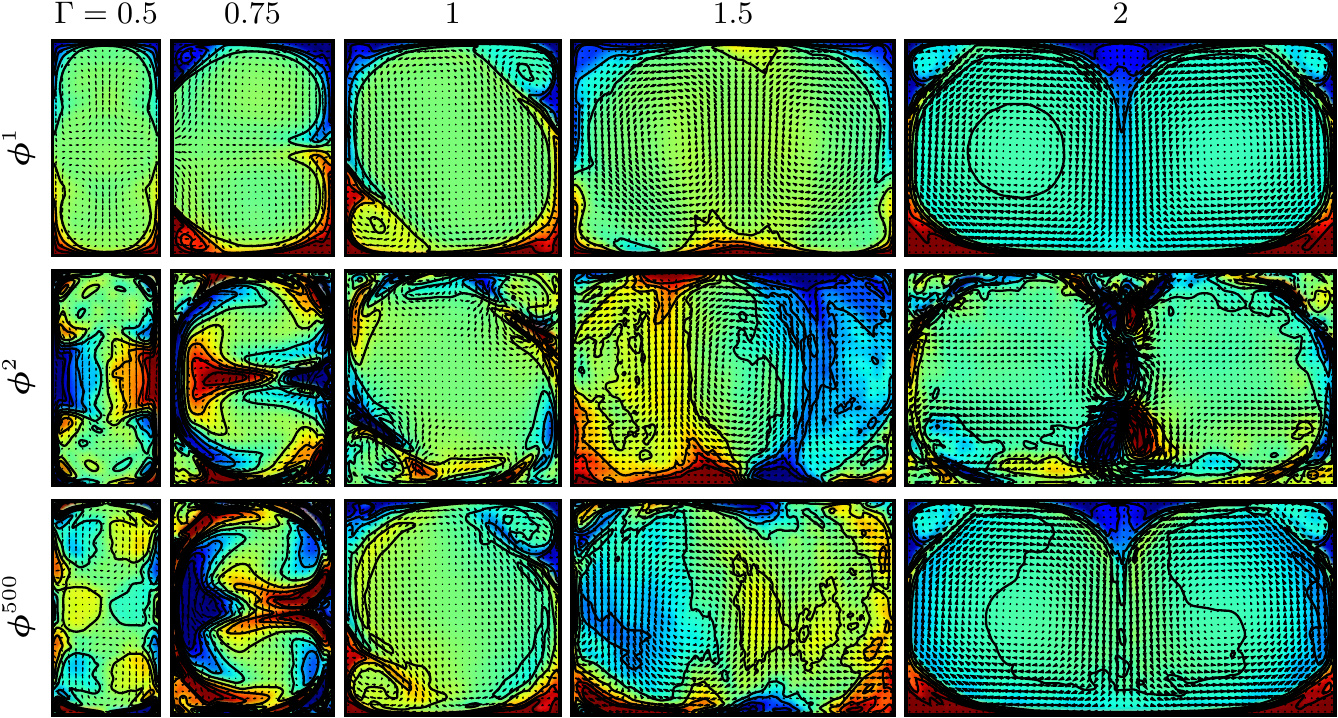}}
	\caption{$\Nu$-modes $\vec{\phi}^1$, $\vec{\phi}^2$ and $\vec{\phi}^{500}$ (i.e., the two most positive and the most negative mode) for five different aspect ratios with $\Ra=10^8$ and $\Pr=1$.\label{fig:gammascan_pods}}
\end{figure}
Although there is a stable large-scale structure for $\Gamma=0.75$, nevertheless the modes following the first one also show a considerable amount of heat transport when compared to $\Gamma\in\{1,2\}$.
We attribute this to the form of the large-scale structure of two vertically stacked rolls:
As the first mode (cf. figure~\ref{fig:gammascan_pods}) does not show vertical structures that extend from one plate to the other, the oscillation of the interface between the rolls that is encoded in the second and third mode becomes important for the heat transport.
On the other hand, for $\Gamma\in\{1,2\}$ the higher modes mainly deform the corner rolls or, for $\Gamma=2$, trigger the oscillation of the cold plume, which has less of a significance for the heat transport.

For $\Gamma=0.5$ the modes are again horizontally symmetric as the convection features regular reversals, and the reversals seem to be already described by the first three and the last mode, as indicated by the eigenvalues in table~\ref{tab:gammascan_relative_heattransport}; this is comparable to the $\Ra=10^7$-case of section~\ref{sec:results_gammascan}, although there, only three modes seem to be involved in the reversal process.
The POD modes for $\Gamma=1.5$ do not show a clear large-scale structure, as the convection is not in a determined roll state but erratic and incoherent.
Therefore the snapshots are less correlated, and thus, more modes are needed for a description of the snapshots.
This becomes especially clear from the percentages of the eigenvalues given in table~\ref{tab:gammascan_relative_heattransport}, as they show the flattest spectrum of all the cases we analysed.

\section{Summary and Outlook}
% \begin{itemize}
% 	\item We extended the POD technique to measure the physically meaningful convective heat transport in terms of the Nusselt number.
% 	\item Modes can have a positive or negative heat transport
% 	\item Projection onto the positive modes show the structures of strong heat transport; projection onto negative modes show structures that hinder the heat transport.
% 	\item New modes perform better with fewer modes in describing the heat transport
% 	\item Outlook 1: Stochastical model: Estimate statistics (e.g., PDF and spectrum) of the time series of the modes; generate random time series of the most important modes to generate artificial data sets
% 	\item Outlook 2: Study the modes / projections with negative heat transport, i.e. the structures that hinder the heat transport. Use this information to actively enhance heat transport in experiment by suppressing these negative structures.
% \end{itemize}
In this paper we extended the technique of proper orthogonal decomposition to measure the convective heat transport in terms of the Nusselt number, thus adapting it to the Rayleigh--Bénard system so that it yields physically more meaningful results.
These so-called $\Nu$-modes give an orthonormal basis set that we used to analyse lower-dimensional projections of data sets generated from direct numerical simulations.

We first benchmarked the proposed method by comparing it to the usual description in terms of the $L^2$-norm or generalised energy, i.e. sum of kinetic energy and temperature variance (dubbed \enquote{$E$-modes}).
We demonstrated that the $\Nu$-modes outperform the $E$-modes in capturing the convective heat transport of the data set, i.e. compared to the $E$-modes, fewer $\Nu$-modes suffice to achieve the same level of $\Nuc$.
Also, lower-dimensional projections onto $\Nu$-modes give a far better reconstruction of the time series of the convective heat transport with fewer modes.
By calculating the PDF of the local $\Nuc(\vec{x})$, we have seen that the $\Nu$-modes can distinguish between positive and negative events of heat transport, and in fact the modes can be grouped into positive and negative ones, while the $E$-modes are insensitive to the heat transport.

We then proceeded to apply the new method to three-dimensional convection in a cylindrical vessel, where the $\Nu$-modes also outperformed the $E$-modes, albeit to a lesser extend.
We attribute this to the observation that in three dimensions, events of negative convective heat transport occur less often.
In the end we analysed the $\Nu$-modes and their spectrum of two-dimensional convection with varying Rayleigh number and with varying aspect ratio.
We found that for erratic flows that do not display a stable large-scale circulation and are instead governed by reversals or incoherent flow structures without a clear roll pattern, the spectrum of the Nusselt modes falls off less steeply.
This indicates that more (positive as well as negative) modes are active in unstructured flows when compared to convection with a stable large-scale circulation.

We want to point out that we deliberately did not take into account the reflectional symmetries the Rayleigh--Bénard system in general has, because these are obviously broken when considering data sets with a large-scale circulation in a preferred direction.
Also, we want to remark that we are aware that the \enquote{benchmark} of describing the convective heat transport is biased, as conversely the $E$-modes outperform the $\Nu$-modes in e.g. reconstructing the time series of the generalised energy.
Nevertheless, the argument still holds that the Nusselt number is a physically relevant quantity, in contrast to the generalised energy.

As an outlook, the $\Nu$-modes could be used to stochastically generate artificial data sets of arbitrary size:
The temporal statistics and dynamics (represented by e.g. first moments and temporal power spectra of the prefactors) and the importance (represented by eigenvalues) of the modes is obtained by examining data sets of direct numerical simulation.
With this knowledge, arbitrarily long time series of random prefactors that obey the same statistics can be generated.
The optimality of the $\Nu$-modes with respect to the heat transport then suggests that the heat transport by the artificially generated data set is close the one of the direct numerical simulation, i.e. close to the real physics.
The temporal behaviour of the modes might also be obtained from a dynamic mode decomposition \citep{schmid10jfm} that also gives the frequency of each mode; it might be worthwhile to extend this technique to also yield a description in terms of the heat transport.

A distant goal could be to investigate the negative modes more closely, as projections on these modes give the structures that reduce the convective heat transport.
Their examination might then suggest measures to actively suppress the structures of negative heat transport and thus enhance the total heat transport, e.g. by equipping the fluid vessel with appropriate barriers or wall structures.
This suggestion remains as a task for future efforts, though.

The author acknowledges the valuable comments and suggestions by Detlef Lohse and Michael Wilczek as well as Richard J. A. M. Stevens for kindly providing the cylindrical data.
Funding was provided by the DFG (Deutsche Forschungsgemeinschaft) under grant FR~1003/10-1.

\bibliographystyle{jfm}
\bibliography{references}

\end{document}